\documentclass[prx,aps,twocolumn,amsmath,amssymb]{revtex4-1}
\pdfoutput=1
\usepackage{hyperref}
\usepackage{graphicx}
\usepackage{bm}
\usepackage{color}

\usepackage{lipsum}
\usepackage{wrapfig}

\def\cal#1{\mathcal{#1}}
\def\eqq#1{Eq.~(\ref{#1})}
\def\eq#1{(\ref{#1})}
\def\av#1{\langle #1 \rangle}
\def\f#1{Fig.~\ref{#1}}

\def\sec#1{Section~\ref{#1}}
\def\app#1{Appendix~\ref{#1}}
\def\c#1{~\cite{#1}}

\def\x{{\bm x}}
\def\s{{\bm s}}
\def\a{{\bm a}}
\def\as{{\bm a_s}}
\def\A{{\bm A}}

\def\s{s}
\def\a{a}
\def\as{a_s}
\def\A{A}

\def\gb{\gamma_{\rm b}}
\def\gr{\gamma_{\rm r}}

\def\lb{\lambda_{\rm b}}
\def\lr{\lambda_{\rm r}}
\def\lbs{\lambda_{{\rm b},s}}
\def\lrs{\lambda_{{\rm r},s}}

\def\e{{\rm e}}

\def\beq{\begin{equation}}
\def\eeq{\end{equation}}
\def\bea{\begin{eqnarray}}
\def\eea{\end{eqnarray}}

\begin{document}

\title{Rare behavior of growth processes via umbrella sampling of trajectories}
\author{Katherine Klymko$^1$}
\author{Phillip L. Geissler$^1$}
\author{Juan P. Garrahan$^2$}
\author{Stephen Whitelam$^3$}\email{{\tt swhitelam@lbl.gov}}

\affiliation{$^1$Department of Chemistry, University of California at Berkeley, Berkeley, CA 94720, USA \\
$^2$School of Physics and Astronomy, University of Nottingham, Nottingham NG7 2RD, UK\\
$^3$Molecular Foundry, Lawrence Berkeley National Laboratory, 1 Cyclotron Road, Berkeley, CA 94720, USA}

\begin{abstract}
We compute probability distributions of trajectory observables for reversible and irreversible growth processes. These results reveal a correspondence between reversible and irreversible processes, at particular points in parameter space, in terms of their typical {\em and} atypical trajectories. Thus key features of growth processes can be insensitive to the precise form of the rate constants used to generate them, recalling the insensitivity to microscopic details of certain equilibrium behavior. We obtained these results using a sampling method, inspired by the ``$s$-ensemble'' large-deviation formalism, that amounts to umbrella sampling in trajectory space. The method is a simple variant of existing approaches, and applies to ensembles of trajectories controlled by the total number of events. It can be used to determine large-deviation rate functions for trajectory observables in or out of equilibrium.
\end{abstract}
\maketitle

\section{Introduction}

Many growth and self-assembly processes result in patterns or structures that are not in thermal equilibrium. \c{kremer1978multi,stauffer1976kinetic,schmelzer2000reconciling,scarlett2010computational,kim2008probing,sanz2007evidence,peters2009competing,whitelam2014self}. 
Understanding both the typical and rare outcomes of these processes is important for predicting and controlling the structures that result. Simulation models of these processes can provide insight here, as long as one can thoroughly sample models' dynamic trajectories out of equilibrium. Typical trajectories can be sampled by direct simulation, but rare trajectories must be generated using enhanced sampling methods. Existing methods for sampling nonequilibrium probability distributions include forward-flux sampling\c{allen2009forward}, transition-path sampling\c{bolhuis2002transition}, the use of driven or auxiliary dynamics\c{bucklew1990large,chetrite2014nonequilibrium,chetrite2015variational}, and variations of diffusion Monte Carlo\c{giardina2011simulating,giardina2006direct,nemoto2016population,lecomte2007numerical,nemoto2016iterative}.
These methods have proven powerful for a range of non-equilibrium systems. 

In this paper we describe a simple method for efficiently sampling rare trajectories of stochastic processes. Each element of the method is well-known in the mathematical physics literature, but the particular combination we use is not a standard protocol and so we describe it in some detail. The method can be applied to models that can be simulated using a continuous-time Monte Carlo scheme, does not require detailed balance, or that transitions possess a reverse counterpart with nonzero rate. It is therefore well-suited to the study of growth processes, whose dynamics may lack detailed balance, or which may be ``irreversible'' in the sense that particles, once added, cannot be removed. We use the method to identify a connection between the trajectory ensembles of reversible and irreversible models of growth\c{morris2014growth,klymko2016similarity}. These models describe the evolution of a mean-field structure composed of two particle types. They can be regarded as models of the irreversible growth of a bacterial colony or a network, and of reversible molecular self-assembly. They display, at the level of ensembles of trajectories, nonequilibrium phase transitions in which the typical outcome of the growth process changes from being enriched in one component type to being an equal mixture of both types. These transitions are similar to nonequilibrium phase transitions seen in lattice models of irreversible\c{eden1961two,ausloos1993magnetic,candia2008magnetic} and reversible growth\c{whitelam2014self}; the mean-field models therefore merit study as possible prototypical examples of distinct types of growth process. Here we show that typical {\em and} rare properties of the trajectory ensembles of these models are similar at certain points in parameter space. This similarity suggests a form of universality, i.e. that qualitative features of certain growth processes do not depend on the precise values of the rates used to generate them.

The method we use here is inspired by the dynamical large-deviation or  
$s$-ensemble formalism\c{touchette2011basic,ruelle2004thermodynamic,garrahan2007dynamical,Lecomte2007,Chetrite2014,touchette2009large,garrahan2009first,hedges2009dynamic}, but does not attempt to sample the $s$-ensemble. Instead, it involves the use of a probability-conserving auxiliary or reference model\c{touchette2009large} in which events that are rare in the original model are made typical, in order to allow efficient sampling of the relevant piece of the probability distribution. This is a strategy that has been widely used to study glassy systems\c{nemoto2016finite,nemoto2016population,jack2013large} and rare events\c{bucklew1990large,chetrite2015variational,touchette2009large,giardina2006direct,maes2008canonical,giardina2011simulating}. The method is a form of importance sampling, and resembles umbrella sampling\c{torrie1977nonphysical} in trajectory space (see also\c{ray2017importance}). In particular, we show that in a stochastic ensemble controlled by the total number of microscopic events, in which time has essentially been ``integrated out'', importance sampling can be carried out using direct simulation alone, without requiring diffusion Monte Carlo (cloning) methods. Cloning is a natural choice for problems, e.g. of glassy dynamics, in which the number of configuration changes per unit time is a key consideration. However, if one is interested in the occurrence of different types of microscopic events, then sampling can sometimes be done in a simple way. Problems of growth and self-assembly can be productively studied in an ensemble that considers only the microscopic processes that have occurred. In this ensemble one can efficiently and thoroughly answer one of the natural questions of this field: What is the nature of the set of structures produced by the growth or self-assembly protocol?

In \sec{rmm} we describe this method. In \sec{sec_coin} we illustrate the method and its interpretation as an umbrella sampling of trajectories using the coin-toss model. In Sections~\ref{sec_eden} and~\ref{sec_rev} we use the method identify a connection between reversible and irreversible growth processes at the level of their typical and atypical trajectories. We conclude in \sec{sec_conc}. 

\section{Sampling of rare events by direct simulation} 
\label{rmm}

Our aim is to sample potentially rare trajectories of a stochastic model. One way to do so is to introduce a reference model whose typical dynamics is in some sense equivalent to the rare dynamics of the original model\c{touchette2009large,chetrite2015variational,giardina2006direct,maes2008canonical,nemoto2016population}. Here we show that a simple version of this approach, in which one neglects explicit consideration of time, is a natural way to study certain growth processes. The method is related to that used in Ref.\c{rohwer2015convergence}, but uses a probability-conserving reference model (rather than the $s$-ensemble).

We start by considering the constant-event-number ensemble in which dynamics is represented as a discrete Markov chain. At each step of the dynamics the system moves from a microstate $C$ to a new microstate $C' \neq C$ with probability $p(C \to C')$, where $\sum_{C'} p(C \to C')=1$. We do {\em not} require that $p(C \to C') \neq 0$ if $p(C' \to C) \neq 0$. We consider the case in which the transition probability 
\beq
p(C \to C') = \frac{W(C \to C')}{R(C)}
\eeq
is derived from the rates $W$ of a continuous-time Markov process; here $R(C) \equiv \sum_{C'} W(C \to C')$ is the exit rate from state $C$. This ensemble is equivalent to the `$x$-ensemble' of\c{budini2014fluctuating} for $x=0$, i.e. when no constraint is placed upon the elapsed time of a trajectory. The ``time'' interval between each step in this ensemble is unity; we consider ``real'' time to have been integrated out~\footnote{In the constant-time ensemble the probability of generating a portion of trajectory in which a jump $C \to C'$ occurs in a time $\Delta t$ is $\e^{-R(C) \Delta t} W(C \to C')$. In the constant-event-number ensemble we track events but not time; the corresponding weight is $\int_0^\infty {\rm d} \Delta t \, \e^{-R(C) \Delta t} W(C \to C')=W(C \to C')/R(C)$.}.

Let $A$ be a dynamic observable, extensive in the number of steps $K$ of the discrete Markov chain. For the models considered here this observable counts the number of binding or unbinding events involving particles of a given type. Our aim is to compute the probability $\rho(\a,K)$ that a trajectory of length $K$ will possess a particular value $\a \equiv {\A}/K$ of the intensive counterpart of the observable $A$. In the large-$K$ limit this computation yields the large-deviation rate function $I(\a)$ for models in which  $\rho(\a,K) \sim \e^{-K I(a)}$\c{touchette2009large}.

The master equation (see \app{master}) in the constant-event-number ensemble for the probability $P(C,A,k)$ of observing dynamic order parameter $A$ and configuration $C$ at step $k$ is
\bea
\label{eom}
P(C,A,k) &=& \sum_{C'} p(C'\to C) \nonumber \\ &\times& P(C',A-\alpha(C' \to C),k-1),
\eea
where $\alpha(C\to C')$ is the change of $A$ upon moving from $C$ to $C'$. The ``$s$-ensemble'' formalism provides a framework for calculating the desired probability $\rho(\a,K)= \sum_C P(C,A,K)$. A central object in this task is the Laplace transform $P_A(C,s,t) \equiv \sum_A \e^{-s A} P(C,A,t)$, which satisfies a master equation 
\bea
\label{eom2}
P_A(C,s,k) = \sum_{C'} p_s(C'\to C) P_A(C',s,k-1), \hspace{0cm}
\eea
with transition ``probability''
\beq
p_s(C \to C') \equiv \frac{\e^{-s \alpha(C \to C')} W(C\to C')}{R(C)}.
\eeq
Note that the ratio of transition probabilities of the original model and $s$-ensemble, $p(C \to C')/p_s(C \to C') = \e^{s \alpha(C \to C')}$, depends in general on the departure state $C$ and the arrival state $C'$. Thus, starting from state $C$, trajectories of the original model and the $s$-ensemble will in general explore a different set of states. In addition, the transition ``probability'' $p_s$ is not normalized, i.e. $\sum_{C'} p_s(C \to C') \neq 1$, and so special techniques are in general required to determine the $s$-ensemble for a given problem \cite{hedges2009dynamic,giardina2011simulating,Chetrite2014}.

Here we proceed not by attempting to sample the $s$-ensemble directly, but by noting that we can mimic some of the properties of the $s$-ensemble using simulations of a probability-conserving reference model. To see this, note that the probability-conserving reference model 
\beq
\label{ref_model}
W_{\rm ref}(C \to C') = \e^{-s  \alpha(C \to C')} W(C \to C')
\eeq
satisfies, in the constant-event-number ensemble, a master equation 
\bea
P_{\rm ref}(C,A,k) &=& \sum_{C'} p_{\rm ref}(C'\to C) \nonumber \\ 
&\times&  P_{\rm ref}(C',A-\alpha(C' \to C),k-1).
\eea
Here 
\beq
p_{\rm ref}(C \to C') \equiv\frac{\e^{-s \alpha(C \to C')} W(C\to C')}{R_{\rm ref}(C)},
\eeq
and $R_{\rm ref}(C) \equiv \sum_{C'} W_{\rm ref}(C \to C')$. It is then apparent that the ratio of transition probabilities of the $s$-ensemble and reference model, $p_s(C \to C')/p_{\rm ref}(C \to C')=R_{\rm ref}(C)/R(C)$, depends only on the departure state $C$ and not the arrival state $C'$. Trajectories of the reference model can thus be ``reweighted'' by factors of $R_{\rm ref}(C)/R(C)$ (one for every state visited) in order to determine the probability of observables $a$ for trajectories in the $s$-ensemble. From there, we can reweight by a factor $\e^{s K a}$ in order to recover the probability $\rho(\a,K)$ of observables $a$ for trajectories of length $K$ in the original model.

To carry out this reweighting we observe that\c{touchette2009large,chetrite2015variational}
\bea
\label{com_first}
\rho(a,K)&\equiv& \sum_\x{P[\x] \delta{(\A[\x]-K \a)}} \\
\label{com} &\equiv& \sum_\x{P_{\rm ref}[\x] w[\x] \delta{(\A[\x]-K \a)}}.
\eea
Here the sums denote path integrals over all trajectories $\x = \{C_0,C_1,C_2,\dots,C_K\}$. The weights of these integrals, $P[\x]$ and $P_{\rm ref}[\x]$, account for the dynamics of the original and reference models, respectively, and are proportional to $\prod_{k=0}^{K-1} p(C_k \to C_{k+1})$ and $\prod_{k=0}^{K-1} p_{\rm ref}(C_k \to C_{k+1})$ (multiplied by a factor accounting for the initial state). The delta function picks out paths consistent with the total dynamic observable of a trajectory, 
\beq
\label{rw1}
\A[\x]=\sum_{k=0}^{K-1} \alpha(C_k \to C_{k+1}),
\eeq
being equal to $Ka$. The ``reweighting factor'' $w[\x] \equiv P[\x]/P_{\rm ref}[\x]$ is the relative probability of a trajectory $\x$ within the original and reference models. This factor is given by a product of terms of the form $p(C \to C')/p_{\rm ref}(C \to C')$, 
\beq
\label{rw0}
w[\x]\equiv \frac{P[\x]}{P_{\rm ref}[\x]}=\e^{s A[\x]+K q[\x]},
\eeq
where 
\beq
\label{rw2}
q[\x] \equiv K^{-1} \sum_{k=0}^{K-1} \ln \frac{R_{\rm ref}(C_k)}{R(C_k)}.
\eeq
In general the reweighting factor fluctuates from trajectory to trajectory. However, we note that $w[\x]$ in the constant-event-number ensemble depends only upon states visited, and contains no factors of time. Consequently, we have found that the logarithm of the sum of $w[\x]$ over trajectories can be evaluated accurately by direct simulation and cumulant expansion. The same strategy is more problematic in the constant-time ensemble, whose reweighting factor depends upon the (variable) times between jumps (see \app{cen}). There, path-sampling\c{hedges2009dynamic} or cloning\c{giardina2011simulating} techniques have been successfuly used.

To evaluate $\rho(a,K)$ we note that \eqq{com}, with $a=a_s$, is the instruction to take the arithmetic mean of the values $w_j$ of the weight functions of typical trajectories $j$ of the reference model; typical trajectories are those that exhibit typical values $a_s$ of our observable. That is, if we generate ${\cal N}$ trajectories of the reference model, and the trajectories labeled $i=1,2,\dots, {\cal M} \leq {\cal N}$ are typical in this sense, then \eq{com} reads 
\bea
\label{inter}
\rho(a_s,K) &=& \frac{1}{{\cal N}}(w_1 + \cdots + w_{\cal M}) \nonumber \\
&\equiv& \frac{{\cal M}}{{\cal N}}\cdot \frac{1}{{\cal M}} (w_1 + \cdots + w_{\cal M}).
\eea
From \eq{rw0} we have $w_j = \e^{s K a_s} \e^{K q_j}$, and so, upon taking logarithms of \eq{inter}, we have
\bea
\label{inter2}
-K^{-1} \ln \rho(a_s,K) &=& -s a_s \nonumber \\
&-&K^{-1} \ln {\cal M}^{-1} (\e^{K q_1} + \cdots +\e^{K q_{\cal M}}) \nonumber \\
&-&K^{-1} \ln ({\cal M}/{\cal N}).
\eea
We can write \eq{inter2} in a more compact way as
\bea
\label{big}
-K^{-1} \ln \rho(a_s,K) &=& - s a_s -K^{-1} \ln \int {\rm d}q\, P_s(q|a_s) \e^{K q} \nonumber \\ &-& K^{-1} \ln \rho_s(a_s,K).
\eea
Here $a_s$ is a value of $a$ typical~\footnote{`Typical' means a value around which trajectories {\em concentrate}\c{touchette2009large}. For models with one attractor this value is the mean value; for models with multiple attractors there can be multiple typical values of the observable.} of the reference model \eq{ref_model} (note that $a_s$ can be a function of $K$); and $P_s(q|a_s)$ is the probability distribution of $q$ for an ensemble of typical reference-model trajectories. By typical trajectories we mean trajectories that have values $\A[\x] = K\a_s$ (in simulations we consider trajectories with values of $A$ within a small window around this value, and we verify that the precise size of the window does not matter). Normalization is such that $\int {\rm d}q\, P_s(q|a_s)=1$. The quantity $\rho_s(a_s,K)$ is the probability of observing a typical reference-model trajectory (defined in the manner above) in an ensemble of reference-model trajectories. 

\eqq{big} involves no approximations, and is valid for all $K$. For large $K$ the term $K^{-1} \ln \rho_s(a_s,K)$ becomes negligible, because $a_s$ is typical of the reference model and so $\rho_s(a_s,K)$ is of order unity (see e.g.\c{touchette2009large}). In this limit $-K^{-1} \ln \rho(a_s,K)$ becomes the large-deviation rate function $I(a_s)$, for models whose probability distributions $\rho(a,K)$ take large-deviation forms $\rho(a,K) \sim \e^{-K I(a)}$:
\bea
\label{ld1}
I(a_s) &=& - s a_s -K^{-1} \ln \int {\rm d}q\, P_s(q|a_s) \e^{K q}.
\eea
We can also write \eq{ld1} as
\beq
\label{ld2}
I(a_s) =- s a_s -q_s-K^{-1} \ln \int {\rm d}q\, P_s(q|a_s) \e^{K \delta q},
\eeq
where $q_s \equiv \int {\rm d}q\, q \, P_s(q|a_s)$ is the mean value of $q[\x]$ for the ensemble of typical reference-model trajectories (those with $A[\x] = K a_s$), and $\delta q \equiv q -q_s$.

The procedure used to obtain Equations \eq{inter2}--\eq{ld2} involves elements that are well-known in the mathematical physics literature, but the particular combination we use is not standard, and results in a particularly simple simulation scheme, as discussed below. Equations \eq{ref_model} and \eq{com} define the most common choice of importance sampling by exponential change of measure, intended to make an event rare in the original model common in the reference model\c{bucklew1990large,touchette2009large}. The simple change of measure defined by \eqq{ref_model} leads to accurate evaluation of Equations \eq{inter2}--\eq{ld2} in the constant-event-number ensemble. In this ensemble the weight $q[\x]$ contains no factors of time and depends only on states visited, and as a result the integrals in these equations can, for some models, be evaluated straightforwardly, by calculation of only the mean and variance of $P_s(q|a_s)$ (for some models $q$ does not fluctuate at all; see e.g. \f{fig1}). In the constant-time ensemble, by contrast, the weight fluctuates more strongly from trajectory to trajectory (see \sec{cen}), and so additional techniques are required to evaluate probabilities. These techniques include cloning\c{giardina2006direct,giardina2011simulating}, the use of ``optimal'' changes of measure\c{Chetrite2014,chetrite2015variational} that render the reference model close to or equivalent to the $s$-ensemble~\footnote{Such techniques would produce reference models whose rates effectively absorb the integral in Equation \eq{ld1}, so that it does not appear in the weight function.}, and adaptive sampling\c{nemoto2016finite,nemoto2016population}. For the models to which we have applied the present method, none of these techniques is necessary. Finally, equations \eq{inter2}--\eq{ld2} are a direct representation of the probability of the value $a_s$ in the original model, derived heuristically and without appeal to the formalism of large-deviation theory. The structure of these equations resembles a parametric Legendre transform of a scaled cumulant-generating function\c{ellis2007entropy,touchette2009large}, but unlike the Legendre transform these equations can recover non-convex rate functions; see e.g. Section~\ref{sec_rev}.

Thus the combination of methods we use involves a simple change of measure; the constant-event-number ensemble; and direct evaluation of probabilities. The result is a method that is simple to implement, involves only direct simulation of a probability-conserving reference model, and requires no rejection of states or trajectories. 

To evaluate \eqq{big} we simulate the reference model \eqq{ref_model} for a particular choice of $s$. We determine the typical value $a_s$ and the distribution $P_s(q|a_s)$ for the ensemble of reference-model trajectories and, if necessary, the likelihood $\rho_s(a_s,K)$ that a trajectory of the reference model is typical, and insert these values into \eqq{big}. This procedure produces one point $(\as,g(\as,K))$ on the curve $g(\a,K)$. Carrying out the same procedure for different values of $\s$ allows reconstruction of the whole curve $g(\a,K)$. For large $K$ this procedure furnishes the large-deviation rate function $I(a)$.

Numerical evaluation of integrals of the form that appears in \eqq{big} is demanding\c{rohwer2015convergence}, requiring in general a number of trajectories exponential in the trajectory length $K$. However, when the statistics of $P_s(q|a_s)$ is Gaussian or close to it, the logarithm of the integral can be evaluated by low-order cumulant expansion. Evaluation of low-order cumulants requires a number of trajectories that does not scale exponentially with $K$ (see \app{cumulant}). In the constant-event-number ensemble, fluctuations of $q$ are related to fluctuations of occupancies of microstates, and for the models we have encountered such fluctuations are approximately Gaussian (see e.g \f{fig_eden_lattice}) or clipped or skewed Gaussian (see \app{fourstate}). This property allows numerical estimation of the logarithm of the integral without excessive numerical effort. The same procedure would not work in the constant-time ensemble, where the analog of $q$ is drawn from a heavy-tailed distribution (see \app{cen}).

We can use Jensen's inequality to compute a rate-function bound $I_0(a_s) \geq I(a_s)$, where 
\bea
\label{jensen}
I_0(a_s) = - s a_s -q_s
\eea
is the first two terms of \eq{ld2}. Recall that $q_s \equiv \int {\rm d}q\, q \, P_s(q|a_s)$ is the mean value of $q[\x]$ for the ensemble of typical reference-model trajectories, i.e. those with $A[\x] = K a_s$. For some of the growth models considered here the fluctuations of $q[\x]$ vanish in the large-$K$ limit, because the exit-rate ratio $R_{\rm ref}(C)/R(C)$ is fixed by $a$, which is in turn fixed by the delta-function path constraint. In this case the Jensen bound \eq{jensen} is exact. 

For other models (e.g. the lattice model described in Appendix~\ref{lattice}; see \f{fig_eden_lattice}) we have found that $P_s(q|a_s)$ is often Gaussian in $q$ (even when $\rho(a,K)$ is strongly non-Gaussian in $a$), in which case the integral in \eq{ld2} can be evaluated exactly to give
\beq
\label{gauss}
I(a_s) = - s a_s -q_s -\frac{K}{2}\sigma_s^2.
\eeq
Here $\sigma_s^2 \equiv \int {\rm d}q\, q^2 P_s(q|a_s) -q_s^2$ is the variance of $q[\x]$ for the ensemble of typical reference-model trajectories. In \app{fourstate} we present an example in which fluctuations of $q$ are not Gaussian, and one must use more than two cumulants in order to evaluate the integral in \eqq{big}.

For large $K$, evaluation of \eq{big} (or of \eq{jensen} or \eq{gauss} under the appropriate conditions) produces one point $(\as,I(\as))$ on the rate-function curve $I(\a)$. Carrying out the same procedure for different values of $\s$ allows reconstruction of the whole curve $I(a)$. Note that, unlike normal umbrella sampling, we do not require overlapping windows to reconstruct the curve. 
 
\section{An example: tossing coins} 
\label{sec_coin}
\begin{figure*}[t] 
   \centering
   \includegraphics[width=\linewidth]{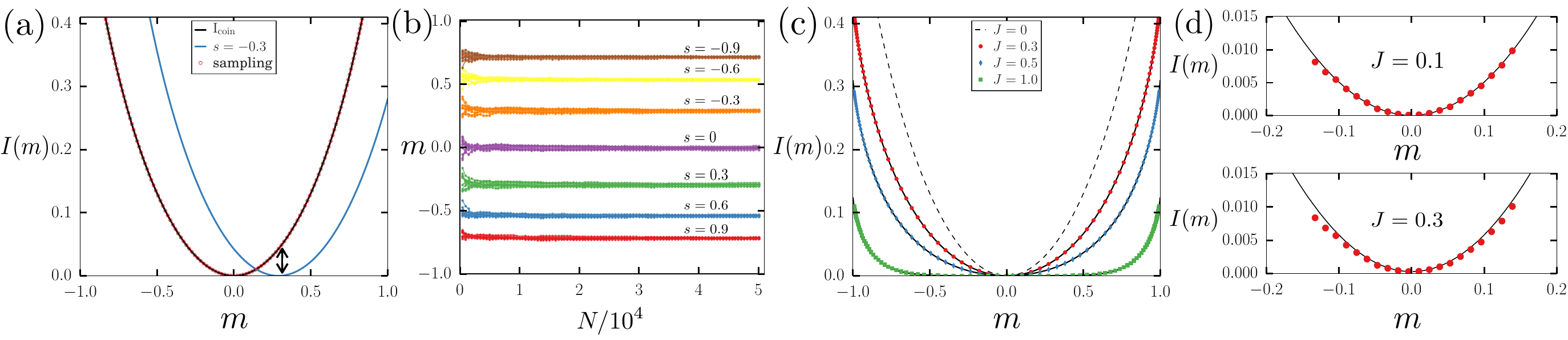} 
   \caption{The reference-model method described in \sec{rmm} can be viewed as umbrella sampling in trajectory space. (a) Large-deviation rate function $I(m)$ for the ``magnetization'' of the coin-toss model, computed numerically (circles) by evaluating \eq{big} for reference-model simulations performed at various values of the bias parameter $s$. These results agree with the known result \eq{ans}. Also shown on the figure is the rate function for the reference model at $s=-0.3$; this can be viewed as an ``umbrella potential'' in trajectory space that concentrates trajectories near a desired region in order parameter space; see also \app{us}. (b) Trajectory ensembles of the coin-toss model and of its reference model at various $s$ illustrate this idea. (c) Rate functions for the mean-field Eden model generated numerically (symbols) or analytically (lines) using the reference model method. (d) Comparison of the reference-model method (solid curve) and direct sampling (circles) validates the reference-model method within the small range of $m$ accessible to direct sampling.}
   \label{fig1}
\end{figure*}

We can illustrate this procedure by applying it to a model of unbiased coin tosses, a classic example of probability theory. To make contact with the growth models considered next we can regard the tossing of a head or a tail as equivalent to the arrival of a ``blue particle'' or a ``red particle'' to a well-mixed system. Let $b$ and $r$ be the number of such particles, and define the extensive magnetization $M \equiv b-r$ and its intensive counterpart $m\equiv M/N$, where $N \equiv b+r$ is the system size (here equal to the total number of events, $K$). We know that the typical `polymorph' will have $m=0$, i.e. equal numbers of red and blue particles, and we can ask~\footnote{This dual formulation of the coin-toss problem illustrates that it is possible to ask meaningful questions of time-dependent processes even without explicit consideration of time. The classic question is to ask how many heads in a certain number of tosses, regardless of how rapidly coins are tossed.}: how likely are we to observe polymorphs enriched in either type of particle?

First recall the standard treatment of this problem. We know from the binomial theorem that the likelihood of $b$ heads (blue particles) in $K=N$ tosses (events) is $\rho(b/N,N) = 2^{-N} \binom {N} {b}$. We can calculate $g_{\rm coin}(m,N) =-N^{-1} \ln \rho(b/N,N)$ using Stirling's formula $x! \approx \sqrt{2 \pi x}(x/\e)^x$ and the result $m=2 b/N-1$:
\bea
\label{ansm1}
g_{\rm coin}(m,N) \approx I_{\rm coin}(m)+ \frac{1}{2N} \ln \left\{ \frac{\pi N}{2} (1-m^2)\right\}, \hspace{0.5cm}
\eea
where
\beq
\label{ans}
I_{\rm coin}(m)=  \frac{1-m}{2} \ln \left(1-m\right)+\frac{1+m}{2}\ln \left( 1+m \right).
 \eeq
To apply the reference-model method of \sec{rmm} we note that in the original model (of unbiased coin tosses) the probability of a head or a tail is identical. In the language of growth, we can say that the arrival rates of blue ($\lambda_{\rm b}$) and red ($\lambda_{\rm r}$) particles are equal. We choose these rates to be $\lambda_{\rm b}= \lambda_{\rm r}=1/2$. We choose to sample the extensive magnetization $M$, which changes by +1 if a blue particle arrives and $-1$ if a red particle arrives. The reference model, \eqq{ref_model}, therefore has $\alpha(C \to C')=+1$ for blue additions and $\alpha(C \to C')=-1$ for red additions. The reference-model rates are then $\lambda_{{\rm b},s}=\frac{1}{2}\e^{-s}$ and $\lambda_{{\rm r},s}=\frac{1}{2}\e^{s}$. 
\label{sec_eden}
\begin{figure*}[t] 
   \centering
   \includegraphics[width=\linewidth]{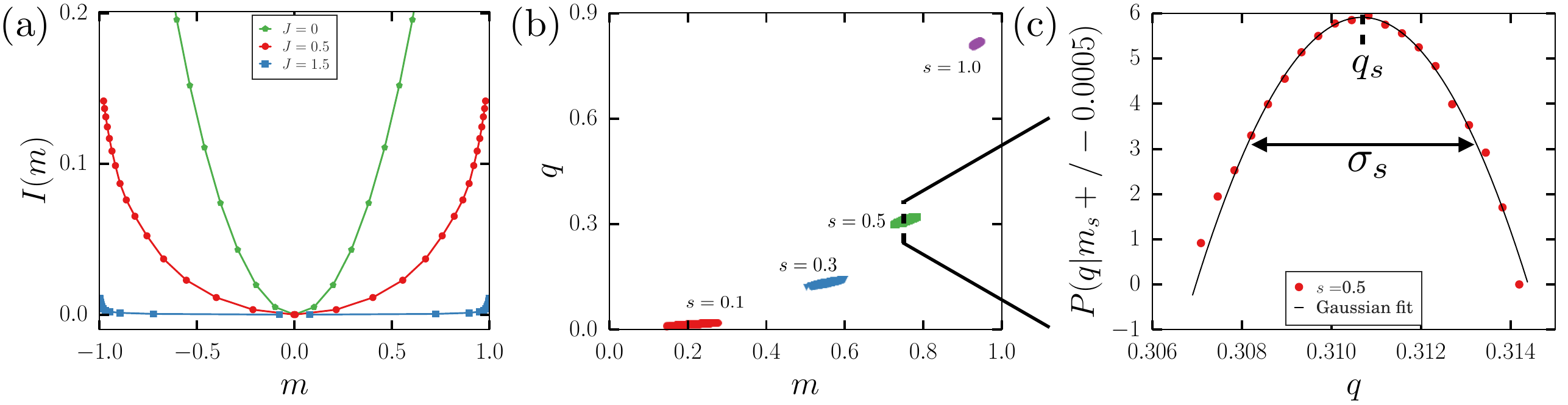} 
   \caption{Demonstration of the reference-model sampling procedure applied to a lattice model. (a) Large-deviation rate function for the lattice magnetic Eden model (see \app{lattice}). Results were produced using direct simulation of a reference model and evaluation of the trajectory-ensemble properties of $m$ and $q$: see panel (b). (c) For the parameters studied the fluctuations of $q$ are Gaussian (even though the fluctuations of $m$ of the original model are strongly non-Gaussian) allowing us to obtain the rate function by evaluation of \eqq{gauss}.}
   \label{fig_eden_lattice}
\end{figure*}

Simulations of the reference model allow evaluation of \eqq{big}. Indeed, for this model, \eqq{big} can also be evaluated analytically. We focus on a large number of events $K=N$, for which the third term in \eq{big} vanishes. Setting $a=m$ and $a_s = m_s$, we can write \eq{big} for the coin-toss model as 
\beq
\label{coin}
I_{\rm coin}(m_s) = - s m_s -N^{-1} \ln \int {\rm d}q\, P_s(q|m_s) \e^{N q}.
\eeq
To evaluate the first term of \eq{coin} note that the typical value of $m$ for the reference model is $m_s=(\lambda_{{\rm b},s}-\lambda_{{\rm r},s})/(\lambda_{{\rm b},s}+\lambda_{{\rm r},s})=- \tanh s$, from which we obtain $s = -\tanh^{-1} m_s$. To evaluate the second term of \eq{coin} note that the quantity $q[\x]$ that appears in the reweighting factor is always $q_s = \ln(R_{\rm ref}(C)/R(C)) =\ln [(\lambda_{{\rm b},s}+\lambda_{{\rm r},s})/(\lambda_{\rm b}+\lambda_{\rm r})]=\ln \cosh s$, and so $P_s(q|m_s) = \delta(q-q_s)$. Thus \eq{coin} reads
\bea
\label{ref_coin} I_{\rm coin}(m_s)&=& - s m_s - q_s \nonumber \\
&=& m_s \tanh^{-1} m_s - \ln \cosh \tanh^{-1} m_s \nonumber \\
&=& \frac{1-m_s}{2} \ln \left(1-m_s\right)+\frac{1+m_s}{2}\ln \left( 1+m_s \right), \hspace{0.8cm}
\eea
giving us one point $(m_s,I_{\rm coin}(m_s))$ on the coin-toss rate-function curve \eq{ans}. Repeating this process for different values of $s$ allows us to generate the complete curve, as shown in \f{fig1}(a). On the figure we show that numerical simulation and analytic evaluation of \eqq{coin} produce identical results, and that both agree with the exact result \eq{ans}. Numerical simulations were done by generating $10^5$ trajectories of the reference model with given $s$, and retaining only those trajectories that were typical in the sense of having values of $m$ within a small window $m_s \pm \epsilon$. We took $\epsilon=0.0005$, which yielded $10^3-10^4$ typical trajectories at each point $s$. Thus with high precision we have evaluated the coin-toss large-deviation rate function from knowledge of the typical behavior of a reference model, without recourse to Stirling's formula or the binomial theorem, or to any of the formal results of large-deviation theory. 


An alternative way of interpreting the method is to view the rate function of the reference model as an ``umbrella potential'' in trajectory space, guiding trajectories to rare parts of the original model's trajectory space. Panels (a) and (b) of \f{fig1} illustrate this interpretation; Appendix~\ref{us} contains the corresponding equations, obtained by straightforward manipulation of \eqq{big}.

\section{An irreversible model of growth} 
We can use the method of \sec{rmm} to quantify the rare behavior of systems in or out of equilibrium, such as the stochastic growth models of Ref.\c{klymko2016similarity}. The first of these is an irreversible model of growth equivalent to a mean-field version of the magnetic Eden model\c{eden1961two,morris2014growth}. Blue and red particles bind (to a mean-field or well-mixed structure) with rates $\lb=\frac{1}{2}\e^{Jm}$ and $\lr=\frac{1}{2}\e^{-Jm}$. Here $J \geq 0$ is a parameter. As in \sec{sec_coin}, the magnetization $m \equiv (b-r)/(b+r)$, where $b$ and $r$ are the number of blue and red particles in the structure. We also define the extensive magnetization $M \equiv b-r$ and system size $N \equiv b+r$. Once bound, particles do not unbind and so $N=K$, the total number of events. This model can also be regarded as a coin-toss model in which the probability of heads or tails depends, via $m$, on the outcome of the prior tosses; when $J=0$ we recover the regular coin-toss model.

The mean-field Eden model possesses a rich phenomenology\c{klymko2016similarity}. It displays a nonequilibrium critical point, at the level of the ensemble of trajectories, when $J=1$. For $J<1$ its trajectories have, in the long-time limit, vanishing mean magnetization; for $J>1$ its trajectories have nonvanishing long-time magnetization. The relaxation time of the mean magnetization of the ensemble of trajectories slows as one approaches the critical point, and becomes logarithmically slow at the critical point; there, direct simulation of the model cannot determine its asymptotic long-time behavior. Thus, although almost trivially simple in construction, the mean-field Eden model possesses phenomenology that is both physically complex and challenging to determine by direct simulation. 

Here we determine this phenomenology by applying the reference-model method. In order to determine the rare behavior of the extensive magnetization $M$ we construct a reference model, \eqq{ref_model}, in which $\alpha(C \to C')=+1$ for blue-particle additions and $\alpha(C \to C')=-1$ for red-particle additions. The reference-model rates are therefore $\lbs=\frac{1}{2}\e^{J m-s}$ and $\lrs=\frac{1}{2}\e^{-Jm+s}$. The properties of this reference model furnish the rare behavior of the original model. For large $K=N$ this behavior can be computed analytically. In this limit the third term of \eq{big} vanishes, and the first two terms can be evaluated analytically. This is so because typical {\em individual} trajectories of the reference model display fluctuations of $m$ that vanish in the long-time limit (even though fluctuations of the trajectory {\em ensemble} of the original model can be large). Typical reference-model trajectories possess, for large $N$, values of the order parameter $m_s$ that satisfy $m_s= \tanh(J m_s -s)$, and each trajectory's value of $q[\x]$ converges to $\ln(R_{\rm ref}(C)/R(C)) =\ln( \cosh(J m_s - s)/\cosh(J m_s))$. Thus \eq{big} can be written
\bea
\label{eden}
I_{\rm Eden}(m_s) &=& m_s (\tanh^{-1} m_s - J m_s) \nonumber \\&-& \ln \frac{\cosh \tanh^{-1} m_s}{\cosh J m_s} \hspace{0.75 cm}\\
&=&  I_{\rm coin}(m_s)- J m_s^2 + \ln \cosh Jm_s.
\eea
Repeating the procedure for a range of values of $s$ generates the rate-function curve
\beq
 \label{eden}
I_{\rm Eden}(m)=  I_{\rm coin}(m)- J m^2 + \ln \cosh Jm.
\eeq
We also calculated \eqq{big} numerically. We generated $10^5$ trajectories of the reference model at each of a range of values of $s$, and identified those trajectories that possessed values of $m$ within a window $m_s \pm \epsilon$ of the typical value $m_s$ (in this case $m_s$ corresponds to the the mean value of $m$ for each set of reference-model trajectories). We took $\epsilon = 10^{-4}$, yielding about $1000$ trajectories for each value of $s$. We verified that the results were insensitive to a value of $\epsilon$ twice as large or twice as small. From this typical ensemble we calculated the terms of \eqq{big}. 
\begin{figure*}[t]  
   \centering
   \includegraphics[width=\linewidth]{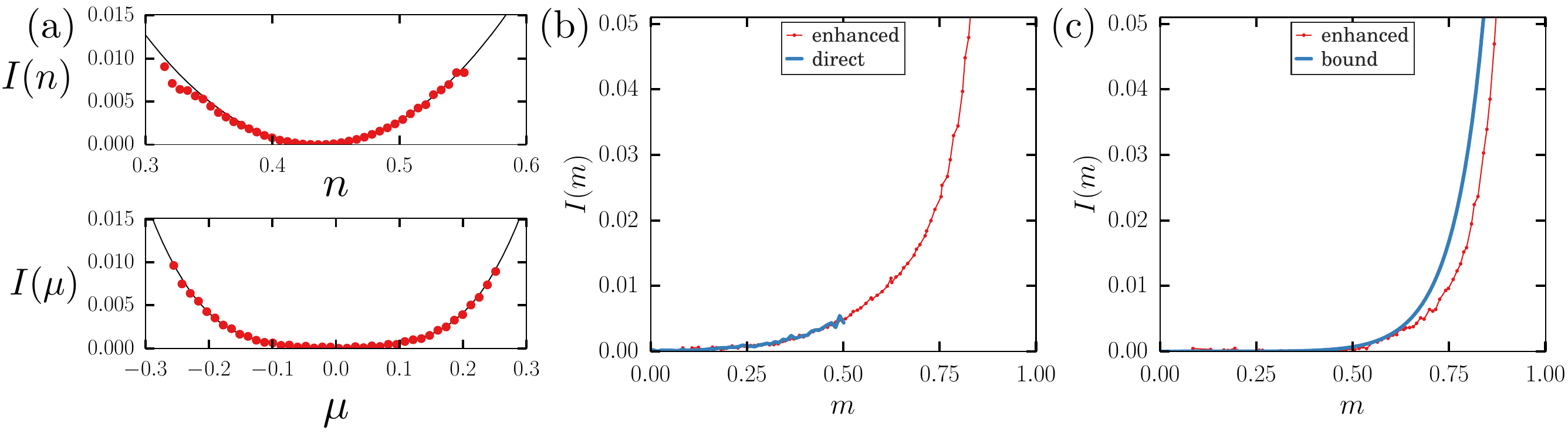} 
   \caption{Demonstration of two-dimensional sampling of the reversible model of growth. Panel (a) shows direct simulation (dots) vs. enhanced sampling (i.e. the reference-model method; lines) for two intensive variables, $n \equiv N/K$ and $\mu \equiv M/K$, whose extensive counterparts $N$ and $M$ we bias. The comparison validates the reference-model method within the small window in which direct sampling is effective. Panel (b) shows the resulting large-deviation rate function for intensive magnetization $m$, together with results from direct simulation, for $J=2.0$ and $c=2.5$. Panel (c) shows enhanced sampling at the critical point $J=c=2$, compared with the bound \eq{rev0}.}
   \label{2_dimensional_sampling}
\end{figure*}
\begin{figure*}[] 
   \centering
   \includegraphics[width=\linewidth]{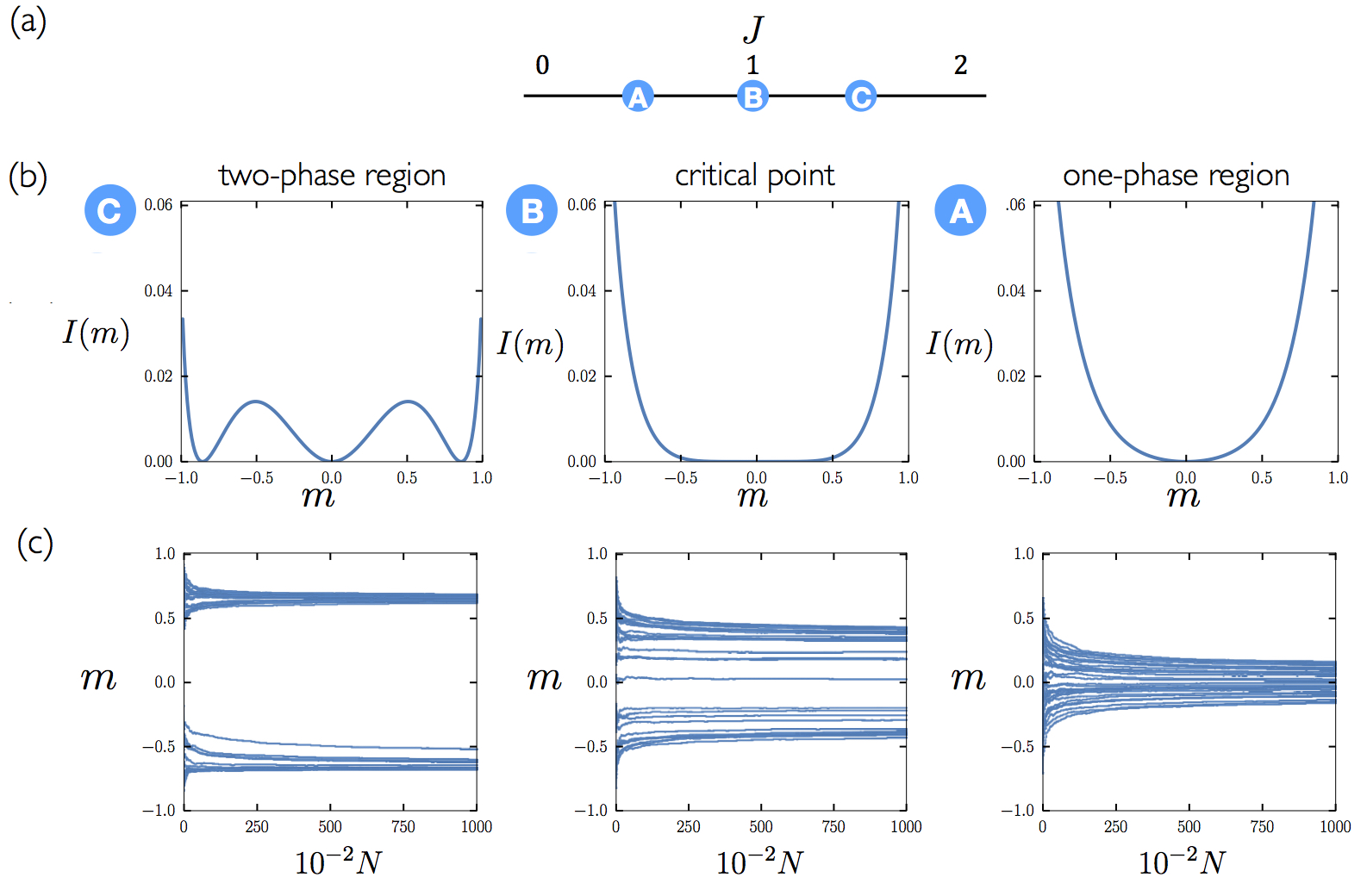} 
   \caption{The irreversible model of growth. (a) Phase diagram; (b) large-deviation rate functions $I(m) = -K^{-1} \ln \rho(m,K)$ for magnetization $m$; (c) a collection of trajectories at the associated phase point. Note that $m$ is a time-integrated observable; each trajectory experiences fluctuations of events (additions of red and blue particles), and the sum of these events results in the behavior shown. In the one-phase region (A) trajectories are of one type, involving the addition (on average) of equal numbers of red and blue particles, and trajectories `concentrate' at the unique minimum of the rate function. At the critical point (B) fluctuations of the trajectory ensemble are anomalously large, in the sense that the rate function is non-quadratic at its unique minimum. In the two-phase region (C) trajectories spontaneously adopt one of two types, involving distinct red-blue addition statistics. The resulting rate function for the time-integrated observable $m$ is non-convex. Moving from A to B to C, the rate function shows behavior qualitatively similar to that of a Landau free energy upon crossing a continuous phase transition.}
   \label{fig_pd1} 
\end{figure*}

In \f{fig1}(c) we show numerical simulation results (symbols) overlaid on \eqq{eden}, for two values of $J$. The Eden model rate function, which is equal to that of the coin-toss model when $J=0$, broadens significantly as one approaches the critical point $J=1$, indicating that fluctuations of the trajectory ensemble are strongly non-Gaussian. On panel (d) we also show the (small) portions of the rate functions at $J=0.1$ and $J=0.3$ accessible by direct simulation. The relaxation time of the model increases so sharply as one approaches the critical point that we could not, from direct simulation, obtain converged results for the rate function close to $J=1$.

The trajectory ensemble of the mean-field Eden model displays strongly non-Gaussian fluctuations for values of $J$ close to the critical point. At the critical point we have a diverging trajectory susceptibility $\chi = N(\av{m^2}-\av{|m|}^2 )\sim N^\theta$, with $\theta \approx 0.8$ measured by direct numerical simulations using trajectories of $N \sim 10^5$\c{klymko2016similarity}. Here the average $\av{\cdot}$ is taken over an ensemble of trajectories. However, this numerical measure is not an accurate determination of the asymptotic value of $\theta$. Relaxation at the critical point is very slow, and trajectories of length $N \sim 10^5$ have not adopted their asymptotic long-time distribution (the magnetization distribution is still evolving). The present results reveal that the true long-time value of the susceptibility exponent is $\theta=2/3$. Maclaurin expansion in $m$ of \eq{eden} gives
\beq
I_{\rm Eden}(m) \approx \frac{(J-1)^2}{2} m^2 + \frac{(1 - J^4)}{12} m^4 + \frac{(3 + 2J^6)}{90}m^6, \hspace{0.5cm}
\eeq
showing that the rate function is very broad (sixth-order in $m$) at the critical point. The corresponding asymptotic trajectory-ensemble susceptibility at the critical point is $\chi^\star = \lim_{N \to \infty} N \av{m^2}_{J=1}$ or
\beq
\chi^\star= \lim_{N \to \infty} N\frac{\int_{-1}^1 {\rm d}m \, m^2 \e^{-w N m^6}}{\int_{-1}^1 {\rm d}m \,  \e^{-w N m^6}} \propto N^{2/3}
\eeq
(here $w=1/18$ is a constant).

We note that the mean-field Eden model and the mean-field Ising model possess similar typical values of magnetization $m$\c{klymko2016similarity}, but we can distinguish them at the level of atypical values of  $m$: the mean-field Ising model has a magnetization rate function quartic in $m$ at the critical point\c{ellis1985entropy}, rather than sixth order, as we found here for the Eden model.

We note also that the reference-model method can be used in a similar manner to determine the large-deviation rate function for magnetization in the lattice-based version of the magnetic Eden model\c{ausloos1993magnetic,candia2008magnetic}. Details are given in \app{lattice}, and the results are shown in \f{fig_eden_lattice}. The procedure used to generate these results is identical to that described so far. A reference model \eq{ref_model} is built to sample the time-extensive magnetization $M$ (and its intensive counterpart $m=M/N$), and direct simulation of the reference model at particular values of $s$ furnishes typical values $m_s$ and an associated distribution of values of the re-weighting quantity $q[\x]$; see \f{fig_eden_lattice}(b). For the parameters shown the distributions of $q$ are Gaussian, in which case \eqq{big} reduces to \eqq{gauss}. We computed the values of $q_s$ and $\sigma_s^2$ (see \f{fig_eden_lattice}(c)), and inserted these into \eqq{gauss}. The result is shown in \f{fig_eden_lattice}(a). 

\section{A reversible model of growth} 
\label{sec_rev}

\begin{figure*}[] 
   \centering
   \includegraphics[width=\linewidth]{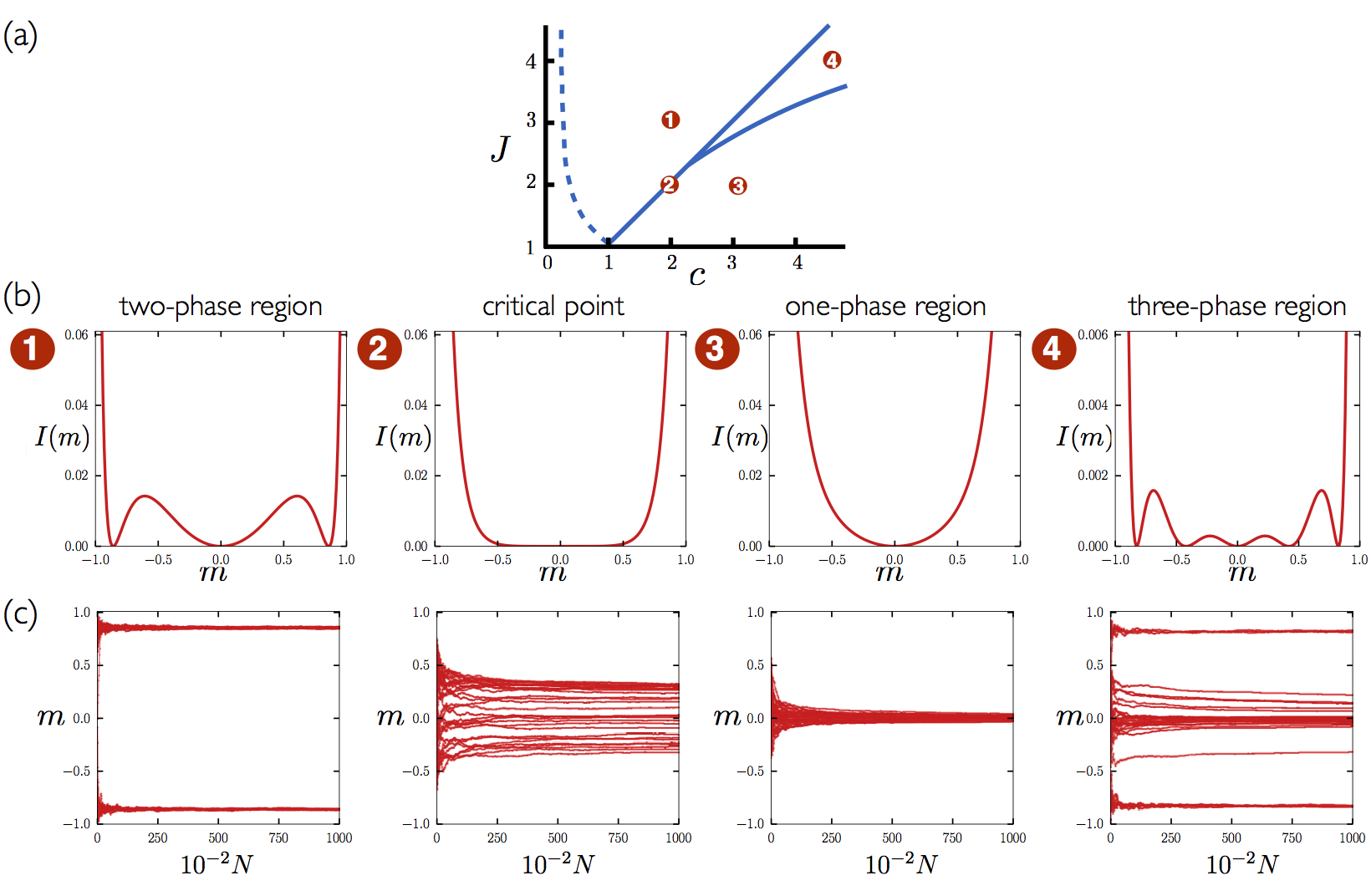} 
   \caption{The reversible model of growth. (a) Phase diagram; (b) large-deviation rate functions for magnetization $m$; (c) a collection of trajectories at the associated phase point. Compare \f{fig_pd1} for the irreversible model. Here the trajectory ensemble undergoes first-order and second-order phase transitions; the three-phase region (point 4) has no counterpart in the irreversible model.}
   \label{fig_pd2} 
\end{figure*}

The other model we consider is the `reversible' model of growth of Ref.\c{klymko2016similarity}, which allows for addition and removal of particles. Reversible and irreversible models of growth are clearly distinct in that only the former can possess an equilibrium. However, we found that at certain points in parameter space the properties of trajectory ensembles of these reversible and irreversible models were identical at the level of typical trajectories: both display a nonequilibrium critical point at which the trajectory ensemble exhibits a diverging susceptibility. Here we use the reference-model method to show that the controlling exponent is the same for the two models, thereby demonstrating a correspondence between these processes at the level of their typical {\em and} atypical trajectories.

We have focused on the behavior of the intensive magnetization $m=M/N$, where $N$ is system size. For an irreversible model of growth the extensive magnetization $M$ can fluctuate, but the system size $N$ is always equal to the length $K$ of a trajectory. For a reversible model the system size is also a fluctuating quantity. As illustrated in \app{toy}, sampling the ratio $M/N$ requires introduction of fields conjugate to both extensive parameters. To do so we note that the reference-model method can be straightforwardly generalized to sample a collection $\bm{A} =(A_1,\dots ,A_L)$ of dynamic order parameters. We introduce a reference model
\beq
\label{ref_model_gen}
W_{\rm ref}(C \to C') = \e^{-\bm{s} \cdot  \bm{\alpha}(C \to C')} W(C \to C'),
\eeq
in which $\bm{s} = (s_1,\dots,s_L)$ is a vector of control variables, and $\bm{\alpha}(C \to C')$ is the change of the vector $\bm{A}$ upon moving from $C$ to $C'$. Proceeding as before gives one point on the curve $g(\bm{a},K) \equiv -K^{-1} \ln \rho(\bm{a},K)$ of the observables $\bm{a} = \bm{A}/K$ of an original model:
\bea
\label{big_gen}
g(\bm{a}_s,K) &=& -\bm{s} \cdot \bm{a}_{\bm{s}} -K^{-1} \ln \int {\rm d}q\, P_{\bm{s}}(q|\bm{a}_{\bm{s}}) \e^{K q} \nonumber \\ &-& K^{-1} \ln \rho_s(\bm{a}_{\bm{s}},K).
\eea
Here $P_{\bm{s}}(q|\bm{a}_s)$ is the normalized probability distribution of the quantity $q[\x] = \sum_k \ln (R_{\rm ref}(C_k)/R(C_k))$ for a set of trajectories of the reference model \eq{ref_model_gen} that are typical, in the sense of having typical values $\bm{a}_s$ of the observables $\bm{A}/K$. As in \eq{gauss}, when the statistics of $q$ are Gaussian, we can write the rate function as
\bea
\label{2d_rate}
I(\bm{a}_{\bm{s}},K)=-\bm{s} \cdot \bm{a}_{\bm{s}}-q_{\bm{s}}-\frac{K}{2}\sigma^2_{\bold{s}}.
\eea
In the reversible model of Ref\c{klymko2016similarity}, blue and red particles add to the system with constant rates $\lb =\lr= c/2$, where $c$ is a notional `solution' concentration. Particles leave the system with non-constant rates $\gb=\frac{1}{2} \e^{-m J} (1+m)$ and $\gr=\frac{1}{2} \e^{m J} (1-m)$ (with $\gr=0$ when $r=0$ and $\gb=0$ when $b=0$), modeling the Arrenhius-like rates with which particles escape from a structure via thermal fluctuations. $J$ sets the energy scale for particle-particle interactions. For $1< J < \sqrt{6}$ the trajectories of this model undergo a continuous nonequilibrium phase transition at $c=J$, similar (at the level of typical trajectories) to that of the mean-field Eden model. Elsewhere in parameter space one observes a region of three-phase coexistence and a line of first-order nonequilibrium phase transitions.

We can sample trajectory ensembles throughout this parameter space by introducing a reference model \eq{ref_model_gen} with $s$ chosen to bias the extensive magnetization $M$ and $s'$ chosen to bias system size $N$: $\lambda_{b,\bm{s}}=\lb \e^{-s-s'}$, $\lambda_{r,\bm{s}}=\lr \e^{s-s'}$, $\gamma_{b,\bm{s}}=\gb \e^{s+s'}$, $\gamma_{r,\bm{s}}=\gr \e^{-s+s'}$. Straightforward algebra (see \app{add}) yields an analytic bound $I_{\rm rev}^0(m)$ on the large-deviation rate function for intensive magnetization $m$:
\bea
\label{rev0}
I_{\rm rev}^0(m)=s_0 m \frac{\Gamma_-(s_0)}{\Gamma_+(s_0)}- \ln \frac{\Gamma_+(s_0)}{\Gamma_+(0)},
\eea
where $\Gamma_\pm(s) \equiv c \cosh s \pm \cosh(s - J m) \pm m \sinh(s - J m)$, and 
\beq
s_0 \equiv - \tanh^{-1}\frac{c m - \sinh(J m) + m^2 \sinh(J m)}{
 c + \cosh(J m) - m^2 \cosh(J m)}.
 \eeq
We also calculated the exact rate function $I(m)$ numerically, by generating approximately $10^5$ trajectories of the reference model with given values of $\bm{s}$ and retaining only those trajectories that were typical in the sense of having values of $n$ and $\mu$ within a small window $\mu_{s_1} \pm \epsilon$ and $n_{s_2} \pm \epsilon$. We took $\epsilon=0.001$, which yielded $10^2-10^3$ typical trajectories at each point $\bm{s}$. This procedure generated a series of points on the surface \eq{2d_rate}; from this surface we can reconstruct $I(m)$ (see \app{toy}, \app{add}). In \f{2_dimensional_sampling} we show how this procedure yields magnetization large-deviation functions for the reversible model at two points in parameter space. At the critical point (panel (c)) we compare the rate function with the analytic bound \eq{rev0}). Expansion of \eq{rev0} reveals the critical rate function to be sixth-order in $m$, just like the Eden model, and so the exponent $\theta$ controlling the divergence of the trajectory susceptibility, $\chi = N(\av{m^2}-\av{|m|}^2 )\sim N^\theta$, is again $2/3$. Thus there exists a correspondence between irreversible and reversible models, at these points in parameter space, in terms of the exponent controlling trajectory fluctuations.
 
In \f{fig_pd1} and \f{fig_pd2} we show trajectories and rate functions for the Eden (irreversible) model and the reversible model throughout the models' dynamic phase diagrams. In the magnetized (multiple-phase) regions of the parameter space the rate functions acquire multiple minima, reflecting the coexistence of distinct types of trajectory. Here the probability distributions of $a$ are multi-model, and so the underlying rate function is non-convex; see e.g. Fig. 6 of\c{touchette2009large}. Other models displaying non-convex rate functions include the Curie-Weiss Ising model below its critical temperature; see e.g. Fig. 1(b) of Ref.\c{paga2017large}. These trajectory types are controlled by stable fixed points in phase space; for the reversible model we have up to three coexisting stable fixed points. Note that these models also possess {\em unstable} fixed points that also show up as {\em minima} (rather than maxima) in the rate functions. This is so because the small-$N$ and large-$N$ behavior of these models is different. Fluctuations of $m$ are large when $N$ is small and small when $N$ is large (see Ref.\c{morris2014growth}): for large $N$, evolution corresponds to Langevin dynamics in the limit of no noise, where the distinction between a stable and an unstable fixed point vanishes. The rate function, obtained in the limit of long trajectories, essentially ignores small-$N$ behavior. For instance, at point (C) in \f{fig_pd1}(a), computing \eqq{big} for small $K$ (small $N$) produces a maximum near $m=0$, while for large $K$ it produces a minimum [\f{fig_pd1}(b)]. One does not see trajectories at the unstable fixed points if one does direct simulation [\f{fig_pd1}(c)], because fluctuations at early times force trajectories to the stable fixed points. Such behavior is qualitatively similar to nucleation, where the presence of phase-space barriers prevents unbiased trajectories from exploring certain regions of phase space.
\section{Conclusions} 
\label{sec_conc}

We have shown that a simple method of rare-event sampling, motivated by the $s$-ensemble formalism and akin to an umbrella sampling of trajectories, can be used to sample the trajectory ensembles of models of reversible and irreversible growth. Of physical significance, our results reveal that key features of trajectory ensembles of certain growth processes are insensitive to the rates of their underlying microscopic processes. This fact suggests the existence of universal features of growth. Of technical significance, the models considered possess a complex phenomenology and are difficult to simulate directly, but they can be analyzed straightforwardly by the sampling method we describe. The method is closely related to a set of sampling methods that have seen wide application in complex systems\c{nemoto2016finite,nemoto2016population,jack2013large,giardina2011simulating}, but differs from those methods in some of its details: it uses the constant-event-number ensemble rather than the constant-time ensemble; the reference model does not attempt to sample the $s$-ensemble directly; and it involves only direct (rejection-free) simulation. The method is simple to implement, natural for certain types of growth process, and can in principle be applied to a wide variety of stochastic processes.
\section{Acknowledgements}
We thank Hugo Touchette, Dibyendu Mandal, and Robert L. Jack for discussions. This work was done as a User project at the Molecular Foundry at Lawrence Berkeley National Laboratory, supported by the Office of Science, Office of Basic Energy Sciences, of the U.S. Department of Energy under Contract No. DE-AC02--05CH11231. K.K. acknowledges support from an NSF Graduate Research Fellowship. J.P.G. was supported by EPSRC Grant No. EP/K01773X/1.


%

\clearpage

\onecolumngrid

\setlength{\parskip}{0.25cm}%
\setlength{\parindent}{0pt}%

\appendix

\section{Master equation in the constant-event-number ensemble}
\label{master}

Consider a Markov process in which $P(C,A,t)$ is the probability of being in microstate $C$ and observing a value $A$ of a dynamic order parameter at time $t$. Let $p(C \to C')$ be the probability per unit time to move from microstate $C'$ to microstate $C$. Then, in time $\Delta t$,
\beq
\label{giggs}
\frac{P(C,A,t) - P(C,A,t-\Delta t)}{\Delta t} = \sum_{C'} p(C' \to C) P(C',A-\alpha(C' \to C),t - \Delta t)-P(C,A,t - \Delta t) \sum_{C'} p(C \to C'),
\eeq
where $\alpha(C' \to C)$ is the change of dynamic order parameter upon moving from $C'$ to $C$. Writing the exit probability per unit time from state $C$ as $r(C) \equiv \sum_{C'} p(C \to C')$ we have
\beq
\label{fergie}
P(C,A,t) = \left[ 1-r(C) \Delta t \right]P(C,A,t - \Delta t)+\sum_{C'} \Delta t \, p(C' \to C) P(C',A-\alpha(C' \to C),t - \Delta t).
\eeq
To pass to the ``$s$-ensemble''  we multiply \eq{fergie} by $\e^{-s A}$ and sum over $A$. Writing 
\beq
P_A(C,s,t) \equiv \sum_{A=-\infty}^\infty \e^{-s A} P(C,A,t)
\eeq
we have
\beq
\label{robson}
P_A(C,s,t) = \left[ 1-r(C) \Delta t \right]P_A(C,s,t-\Delta t) +\sum_{C'} \Delta t \, p(C' \to C) \e^{-s \alpha(C \to C')} P_A(C',s,t - \Delta t).
\eeq
The $s$-ensemble does not conserve probability, as can be verified by summing \eq{robson} over $C$. In this paper we carry out sampling equivalent to that of the $s$-ensemble using a probability-conserving reference model. We focus on the constant-event-number ensemble, meaning the particular case in which the exit probability $r(C) \Delta t$ is unity, and so at each time step the Markov process moves to a new microstate. Such a discrete Markov chain is naturally related to a continuous-time dynamics having rates $W(C \to C')$, i.e.
\beq
p(C \to C') = \frac{W(C \to C')}{\sum_{C'} W(C \to C')} \equiv \frac{W(C \to C')}{R(C)}.
\eeq
We then have $r(C) = 1$ (setting $\Delta t=1$ for simplicity). We can then write \eq{fergie} as
\beq
\label{fergie2}
P(C,A,t) = \sum_{C'} \frac{W(C' \to C)}{R(C')} P(C',A-\alpha(C' \to C),t -1),
\eeq
and \eq{robson} as 
\beq
\label{robson2}
P_A(C,s,t) = \sum_{C'} \frac{W(C' \to C)}{R(C')} \e^{-s \alpha(C \to C')} P_A(C',s,t -1).
\eeq
We then observe that the analog of \eq{fergie2} for a reference dynamics $W_{\rm ref}(C \to C') = \e^{-s \alpha (C \to C')} W(C \to C')$ is
\beq
\label{fergie3}
P_{\rm ref}(C,A,t) = \sum_{C'} \frac{W(C' \to C)}{R_{\rm ref}(C')} \e^{-s \alpha(C' \to C)} P_{\rm ref}(C',A-\alpha(C' \to C),t -1),
\eeq
with $R_{\rm ref}(C) \equiv \sum_{C'} W_{\rm ref}(C \to C')$. The similarity between \eq{robson2} and \eq{fergie3} motivates the reference-model sampling method described in the text. There, time $t$ is renamed $k$.

\section{Reweighting by direct simulation is not feasible in the constant-time ensemble}
\label{cen}

The method described in the main text makes use of direct simulation and reweighting of reference-model trajectories to determine trajectory probabilities for a model of interest. This procedure works in the constant-event-number ensemble, but would be more problematic in the constant-time ensemble. To make this distinction clear, we argue as follows.

The $s$-dependent factors in its rates cause the reference model to exhibit rare values of activity, by directing it to the rare states (and transitions) that give rise to these activities. Thus we have a mechanism for efficiently exploring rare sets of states (and rare transitions). In the constant-event-number (CEN) ensemble, trajectories that visit the same set of states receive the same contribution to the reweighting factor, $\prod_k R_{\rm ref}(C_k)/R(C_k)$, where $k$ labels transitions. Thus, direct simulation is an efficient way of sampling the reweighting factor in this ensemble.

By contrast, in the constant-time (CT) ensemble, each trajectory that visits the same set of states has an infinite number of possible realizations in which jumps between those states are made at different times. \f{fig_cen} illustrates this idea. With direct simulation we have no mechanism for sampling those times efficiently. 

To see this, note that in the constant-time ensemble the probability of generating a portion of trajectory in which a jump $C \to C'$ occurs in a time $\Delta t$ is $\e^{-R(C) \Delta t} W(C \to C')$. The time-dependent piece of the reweighting factor for a single trajectory in the CT ensemble is therefore
\beq
\label{wf}
w = \exp\left(\sum_k[R_{\rm ref}(C_k)-R(C_k)] \Delta t_k\right).
\eeq
Here $k$ labels transitions and the subscript `ref' denotes the reference model. Times $\Delta t_k$ are generated by the reference model from distributions 
\beq
p_k(\Delta t_k)  = R_{\rm ref}(C_k) \exp(-R_{\rm ref}(C_k) \Delta t_k).
\eeq 
The mean $\Delta t_k$ is therefore $1/R_{\rm ref}(C_k)$. However, the weight factor \eq{wf} rewards trajectories whose largest values of $R_{\rm ref}(C_k)-R(C_k)$ are coupled to the largest values of $\Delta t_k$ (subject to $\sum_k \Delta t_k$ being constant), and large values of $\Delta t_k$ are generated with exponentially small probabilities $p_k(\Delta t_k)$. 
\begin{wrapfigure}{L}{0.5\textwidth}
   \centering
   \includegraphics[width=0.5\linewidth]{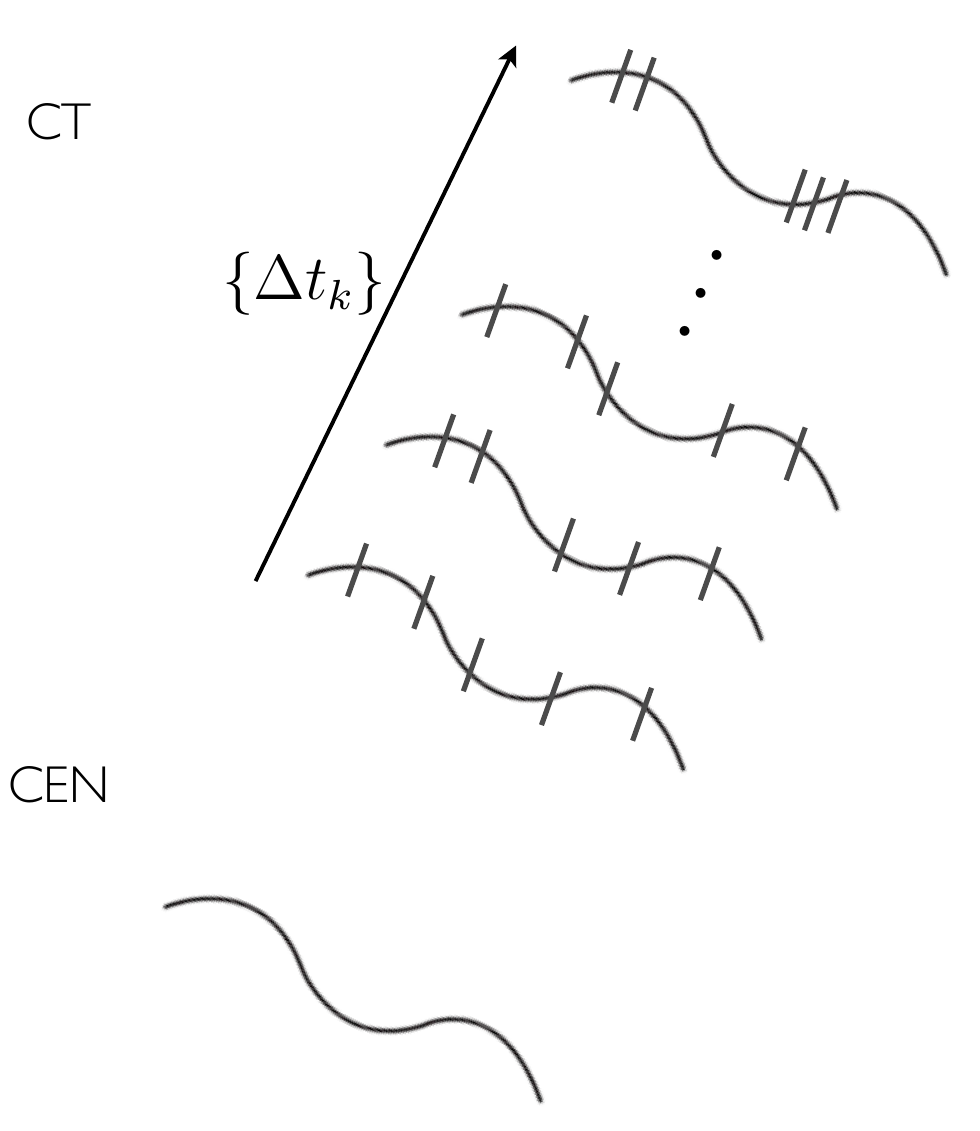} 
   \caption{Trajectories that visit a given set of states in the CT and CEN ensembles are drawn as a polymer. In the CT ensemble each polymer has an infinite number of realizations whose transition times (tick marks) are distributed differently. These realizations carry potentially large weights, but are generated by direct simulation with low probabilities.}
   \label{fig_cen}
\end{wrapfigure}

Consider the simple case of a reference-model trajectory of 2 steps. Let the trajectory jump from state 1 to state 2 after time interval $\Delta t_1$, and from state 2 to state 3 after time interval $\Delta t_2$. Let the escape rates from states 1 and 2 in the original model be unity, i.e. $R(1)=R(2) =1$, and let the escape rates from these states in the reference model be  $R_{\rm ref}(1) = 10$ and $R_{\rm ref}(2) = 2$. Then \eq{wf} reads
\beq
w = \exp(9 \Delta t_1 + \Delta t_2).
\eeq 
The typical value of this factor will occur when $\Delta t_1= 1/10$ and $\Delta t_2 = 1/2$, and is $w=\exp(9/10 + 1/2)= \exp(7/5)$. This value will arise with probability $p_1 \times p_2 \propto \exp(-10/10) \times \exp(-2/2) = \exp(-2)$. The product $p_1 p_2 w$ is $\propto \exp(-3/5)$.

Consider now a reference-model trajectory that makes the same two jumps, with the same total elapsed time, but where $\Delta t_1= 1/2$ and $\Delta t_2 = 1/10$. The likelihood $p_1 p_2$ of generating these time intervals is $\propto \exp(-10/2) \times \exp(-2/10) = \exp(-26/5)$, which is small, but the associated weight is large: $w = \exp(9/2+1/10) = \exp(23/5)$. The product $p_1 p_2  w$ is again $\propto \exp(-3/5)$. Thus the contribution of this rare trajectory to the sum (over all trajectories) of values of $w$ is as significant as the contribution from trajectories with `typical' time distributions. 

With direct simulation of a reference model we cannot efficiently sample these rare trajectories, and so we cannot estimate the weight of a set of trajectories that visit a particular set of states. In the CEN ensemble this problem does not arise: there are no factors of time in the reweighting factor. In effect, by passing from the CT to the CEN ensemble we have performed analytically the integration in the ``time direction'' shown in \f{fig_cen}, yielding a weight 
\beq
\int_0^\infty {\rm d} \Delta t_k \exp\left([R_{\rm ref}(C_k)-R(C_k)] \Delta t_k\right) p_k(\Delta t_k) = \frac{ R_{\rm ref}(C_k)}{R(C_k)}
\eeq
for each state visited in the CEN ensemble (this integration is made possible by relaxing the constraint that each trajectory have the same elapsed time $\sum_k \Delta t_k$). The $s$-bias applied to the rates efficiently directs the reference model to rare states and transitions, so allowing us to sample the reweighting factor in the CEN ensemble.

\section{Cumulant expansion of the reweighting factor}
\label{cumulant}

In main text we evaluate quantities of the form
\begin{equation}
\label{gee}
g \equiv K^{-1} \ln \int {\rm d}q \, \rho(q) e^{Kq},
\end{equation}
where $\rho(q)$ is the probability distribution of $q$ that is generated by a reference dynamics. \eqq{gee} can be written
\beq
\label{expansion}
g=\sum_{n=1}^\infty {1\over n!}K^{n-1}C_n,
\eeq
where $C_n$ is the $n^{\rm th}$ cumulant of $\rho(q)$. The scaling with $K$ of $C_n$ follows from the fact that $\rho(q)$ has a large-deviation form for large $K$ ($q$ is intensive), and is
\beq
C_n \sim K^{1-n}.
\eeq
The mean of $q$ is intensive, $C_1 \sim K^0$, and, as implied by the central limit theorem, its standard deviation scales as $1/\sqrt{K}$, i.e., $C_2 \sim K^{-1}$. Each successively higher-order cumulant decreases in scale by a factor of $K$. Terms in \eq{expansion} are therefore (a priori) of order unity. However, for the models we have studied we have found $\rho(q)$ to be Gaussian, in which case one requires only the first two cumulants, or Gaussian with some skew, in which case one needs three cumulants. Such low-order cumulants can be calculated without excessive numerical effort. Standard considerations indicate that the numerical uncertainty in the calculation of $C_n$, using $N$ trajectories of length $K$, scales as $\sqrt{K^{n-2}/N}$. Thus only $N = {\cal O}(K)$ trajectories are required to evaluate the third cumulant.

\section{Interpretation of the method of \sec{rmm} as an umbrella sampling of trajectories}
\label{us}
In the main text we consider the probability $\rho(a,K)$ of observing a value $A=aK$ of the dynamic order parameter $A[\x]$ for a trajectory of a model (the original model) of length $K$, and we calculate this probability from knowledge of the typical behavior of a reference model. In equations, using the notation of the main text,
\bea
\label{l1} \rho(a,K) &=& \sum_\x P_{\rm ref}[\x] w[\x] \delta{(A[\x]-Ka)} \\
\label{l2} &=&  \rho_s(a,K)\, \e^{s K a} \int {\rm d}q \, P_s(q|a) \e^{K q} \\
\label{l3} &\equiv&\rho_s(a,K) \, \e^{s K a+K \bar{q}}\int {\rm d}q \, P_s(q|a) \e^{K (q-\bar{q})}.
\eea
Here $\bar{q} \equiv \int {\rm d}q \, q\, P_s(q|a)$, and $\rho_s(a,K) =\sum_\x P_{\rm ref}[\x] \delta{(A[\x]-Ka)}$ is the probability that a reference-model trajectory of length $K$ has $A[\x] = Ka$. To motivate the passage from \eq{l1} to \eq{l2}, note that if we generate ${\cal N}$ trajectories of the reference model, and trajectories labeled $i=1,2,\dots,{\cal M} \leq {\cal N}$ have $A[\x] = Ka$, then we can write \eq{l1} as
\bea
\frac{1}{{\cal N}}(w_1+w_2+\cdots+w_{\cal M}) &\equiv& \frac{{\cal M}}{{\cal N}} \frac{1}{{\cal M}}(w_1+w_2+\cdots+w_{\cal M}) \\
\label{l4} &=&\frac{{\cal M}}{{\cal N}} \cdot \e^{s K a}  \cdot \frac{1}{{\cal M}}(\e^{K q_1}+\e^{K q_2}+\cdots+\e^{K q_{\cal M}}).
\eea
The three factors separated by $\cdot$ in \eq{l4} are, in order, the three factors displayed in \eq{l2}. 
 
In the large-$K$ limit we expect $\rho(a,K) \sim \e^{-K I(a)}$ and $\rho_s(a,K) \sim \e^{-K I_{\rm ref}(a)}$; inserting these results into \eq{l3} and taking logarithms gives
\beq
\label{eqa}
I(a) = I_{\rm ref}(a) - s a - \bar{q}  -\frac{1}{K} \ln \int {\rm d}q \, P_s(q|a) \e^{K (q-\bar{q})}.
\eeq
To carry out the procedure described in the main text we sample values $a_s$ of the reference model that are typical in the sense that $I_{\rm ref}(a_s)=0$. In this case we can evaluate
\beq
\label{eqb}
I(a_s) = - s a_s - q_s -\frac{1}{K} \ln \int {\rm d}q \, P_s(q|a_s) \e^{K (q-q_s)},
\eeq
where $q_s\equiv \int {\rm d}q \, q\, P_s(q|a_s)$. The graphical construction shown in Fig. 1 illustrates Equations \eq{eqa} and \eq{eqb}; the rate function of the reference model is used as an ``umbrella potential'' in trajectory space in order to determine the rate function of the original model (in that particular example the quantity $q$ does not fluctuate, and $I(a_s) = - s a_s - q_s$).

\section{4-state model}
\label{fourstate}
Here we demonstrate our method on a 4-state model\c{brown2003single}. For this model the Legendre transform of the largest eigenvalue of the tilted transition matrix can be calculated explicitly to yield $I(a)$\c{touchette2009large}. The extensive activity that we studied was either the total number of horizontal transitions or the number of horizontal transitions on the top branch, as shown in \f{fig_4state}. We show in \f{fig_4state} that the $s$-ensemble and the reference-model ensemble are different, but that the reference-model ensemble trajectories can be reweighted to construct the rate function. We also show that the Gaussian formula \eq{gauss} for the rate function works well when the rates in the model are similar, but that one additional cumulant is required when the rates are numerically very different (in this case fluctuations of the empirical measure from trajectory to trajectory are not Gaussian, because certain states are occupied with close to 100\% probability, and so fluctuations of the weight function are not Gaussian). 

\begin{figure*}[t] 
   \centering
   \includegraphics[width=0.7\linewidth]{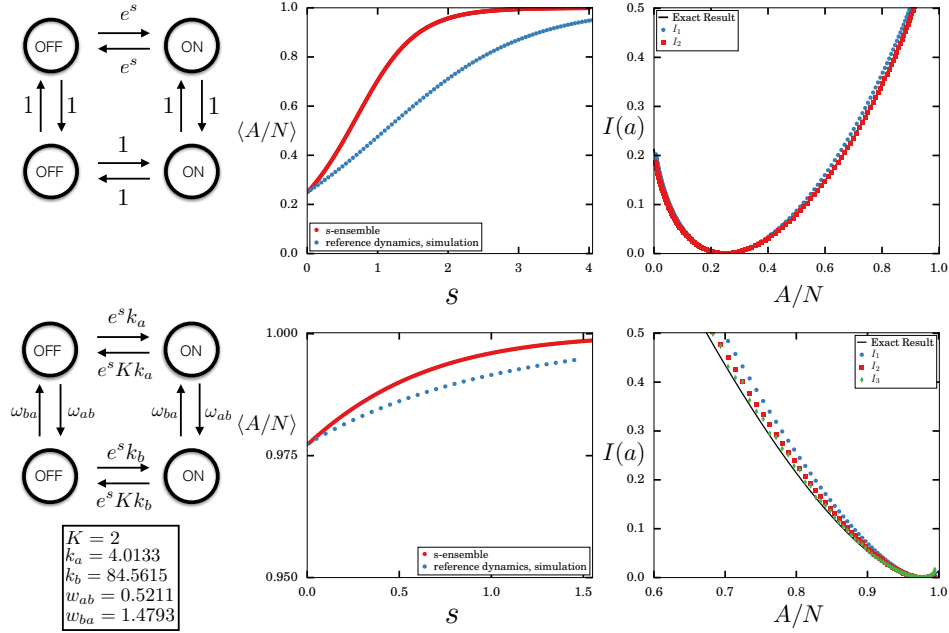} 
   \caption{4-state model with biased horizontal transitions. The leftmost column shows model with the arrows demonstrating the possible transitions with their corresponding rates We show in the middle column the comparison between the average activity in the $s$-ensemble and the reference dynamics. This demonstrates that the reference dynamics is not sampling the $s$-ensemble but we can still reweight to generate the rate function. The third column shows the rate functions. The points shows what come out of enhanced sampling and the black curves the Legendre transform of the cumulant generating function. In the bottom panel, the rates were taken from\c{brown2003single}. Note that when the rates are very similar, the Gaussian formula \eq{gauss} is accurate. When the rates are asymmetric, the third-order cumulant is required.}
   \label{fig_4state}
\end{figure*}

\section{Trajectory distributions of the lattice-based magnetic Eden model}
\label{lattice}
\begin{figure*}[t] 
   \centering
   \includegraphics[width=\linewidth]{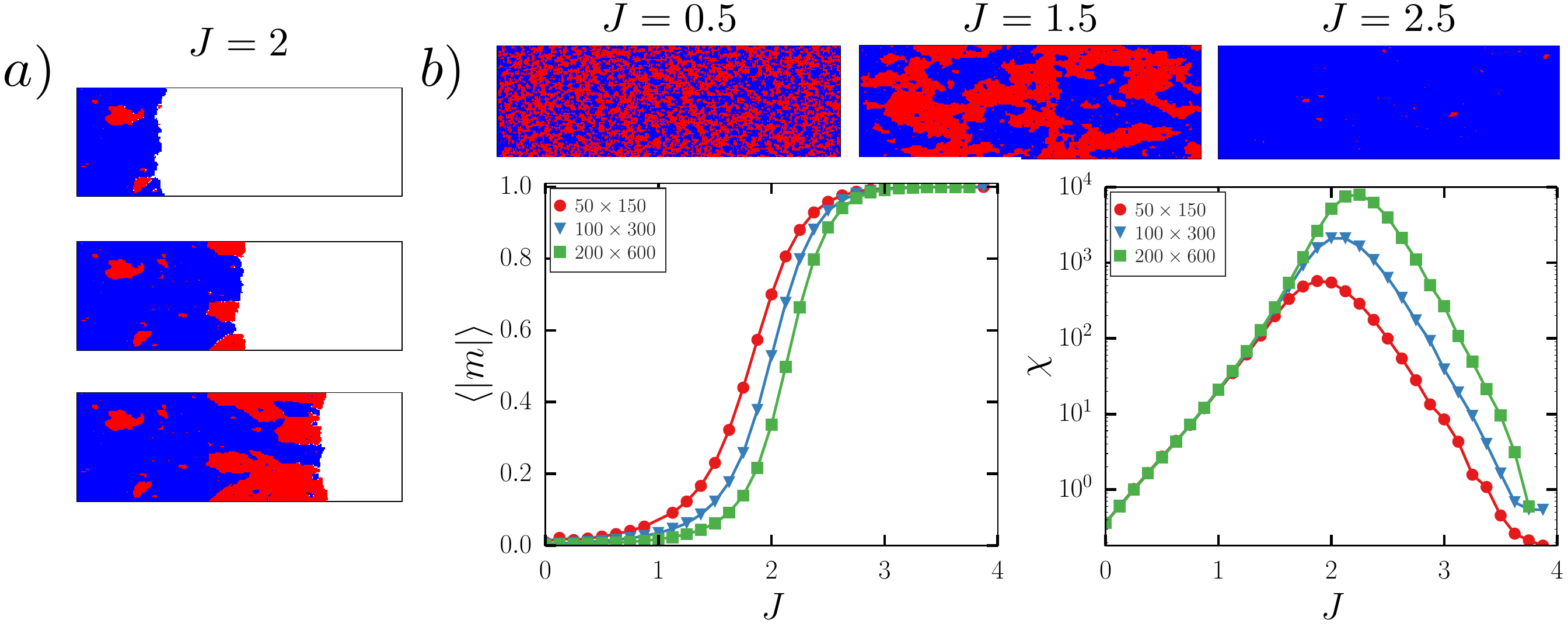} 
   \caption{Phenomenology of the lattice-based magnetic Eden model. (a) Snapshots of a growth trajectory, for $J=2$. (b) Mean magnetization and the trajectory ensemble susceptibility $\chi$ of the magnetization for various values of $J$ indicate the existence of a nonequilibrium phase transition; shown top are snapshots of the products of growth at three values of $J$. In \f{fig_scholes} and \f{fig_eden_lattice} we show the large-deviation rate functions for trajectory ensembles, for three values of $J$.}
   \label{fig_cantona}
\end{figure*}
Here we calculate the large-deviation rate function for magnetization in the lattice model of irreversible growth whose mean-field counterpart is studied in the main text, in order to illustrate the application to a lattice model of the method described in the main text. The lattice model in question is the magnetic Eden model\c{ausloos1993magnetic,candia2008magnetic} in two spatial dimensions. Consider a square lattice on whose sites $i=1,2,\dots, L_x \times L_y$ live spin variables $S_i$. Lattice sites can be empty ($S_i=0$), or occupied by a blue particle ($S_i=1$) or a red particle ($S_i=-1$). We consider rectangular strip geometries of $L_x \times L_y$ lattice sites, where $L_x/L_y=3$. We begin simulations with one line of blue sites at the left-hand edge of the simulation box. 

We simulate in the constant event-number ensemble. Any empty site adjacent to a colored site is {\em active}. We allow any site $i$ that is active to become blue with rate $\lambda_{\rm b}(M_i)= \exp(J M_i)$, and red with rate $\lambda_{\rm r}(M_i)= \exp(-J M_i)$. Here $M_i = \sum_j S_j$, where $j$ runs over the 4 nearest neighbors of $i$. The total escape rate from a microstate is then $R = \sum_{i\, {\rm active}} 2 \cosh (J M_i)$. Thus an active site $i$ will turn red or blue with respective probabilities $\exp(-J M_i)/R$ and $\exp(J M_i)/R$. We carry out one such process, with the appropriate probability, and advance time by one unit. We then update the list of active sites, and repeat until $K$ moves have been carried out.

The phenomenology of this model is shown in \f{fig_cantona}. An example of growth at a fixed value of the parameter $J$ is shown in panel (a); the structure that results contains both red and blue sites. In panel (b) we show the trajectory average $\av{|m|}$ of the absolute magnetization $|m|=N^{-1} \sum_i |S_i|$ of the whole simulation box, as a function of $J$; each trajectory furnishes one value of $|m|$, and each data point shown is obtained by averaging over $10^5$ trajectories. In panel (c) we show trajectory-to-trajectory fluctuations of $|m|$, $\chi = N(\av{m^2} - \av{m}^2)$. The box sizes used were $50 \times 150$, $100 \times 300$, and $200 \times 600$. The behavior shown is indicative of a nonequilibrium phase transition between mixed and demixed arrangements of red and blue sites\c{ausloos1993magnetic,candia2008magnetic}; we observed similar behavior in a lattice model of reversible growth\c{whitelam2014self}. 

\begin{figure}[t] 
   \centering
   \includegraphics[width=0.5\linewidth]{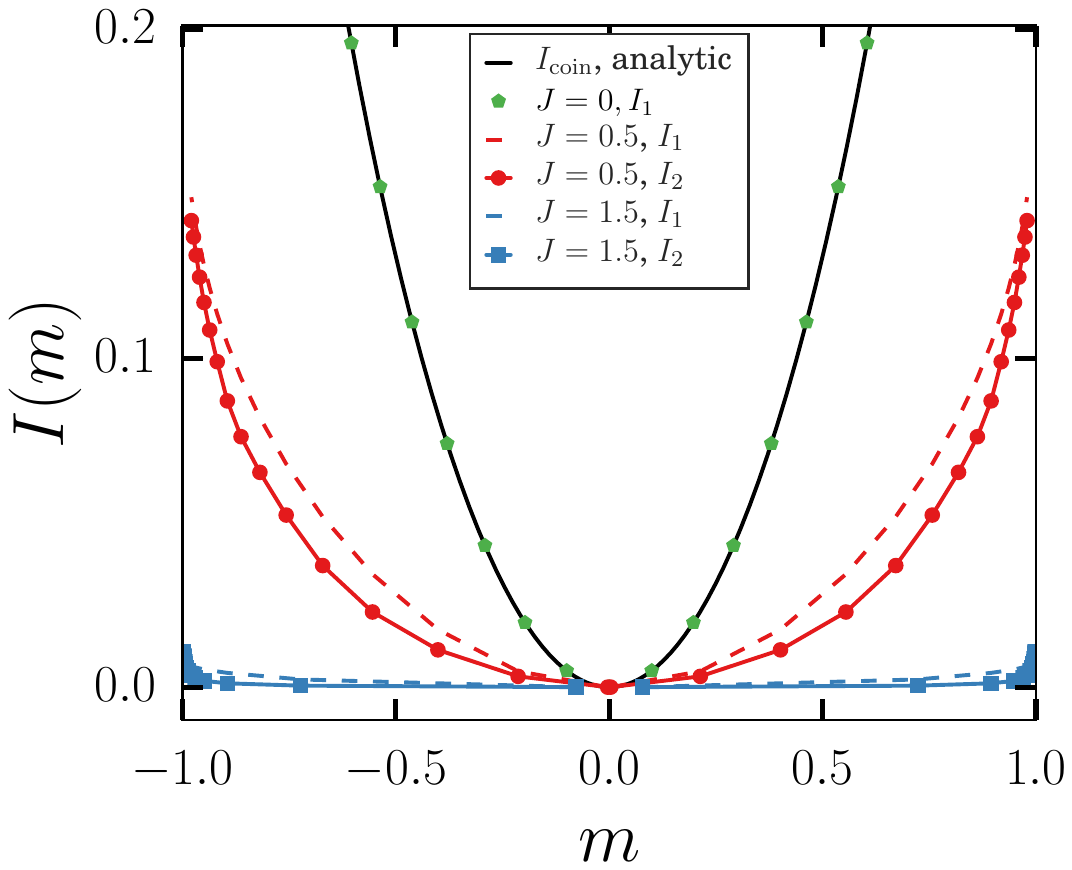} 
   \caption{Rate functions for magnetization in the lattice-based magnetic Eden model, as \f{fig_eden_lattice}, here including the Jensen bound \eq{jensen} (dotted lines). The case $J=0$ coincides with the coin-toss rate function, as it should, because in this case the model describes the creation of red or blue sites with equal likelihood, regardless of the state of neighboring sites. As $J$ increases we observe strongly non-Gaussian fluctuations of the trajectory ensemble, similar to the mean-field Eden model discussed in the main text.}
   \label{fig_scholes}
\end{figure}

To calculate the large-deviation magnetization rate function for the lattice magnetic Eden model we can apply the reference-model procedure described in the main text. We introduce a reference model whose rates are $\lambda_{{\rm b},s}(M_i)= \exp(J M_i-s)$ and $\lambda_{{\rm r},s}(M_i)= \exp(-J M_i+s)$. We simulate the reference model at fixed $s$, and keep track of the quantity $Q_\alpha = \sum_{\rm moves} \ln(R_{\rm ref}/R)$, where $R_{\rm ref} = \sum_{i\, {\rm active}} 2 \cosh (J M_i-s)$. Here $\alpha$ labels trajectories, and $i$ runs over all active lattice sites. 

For each value of $J$ and $s$ considered we generated ${\cal M} = 10^5$ reference-model trajectories. In \f{fig_eden_lattice}(b) we show a scatter plot of values of $(m_\alpha,Q_\alpha)$ for trajectories $\alpha=1,2,\dots,{\cal M}$. We identify `typical' trajectories as those whose resulting structures possess magnetization $m$ within $\pm \epsilon$ of the mean value $m_s$ (the mean is calculated over all $10^5$ trajectories generated). We used a tolerance $\epsilon = 5 \times 10^{-4}$, and verified that the results presented did not change upon halving or doubling $\epsilon$. In \f{fig_eden_lattice}(c) we show that the distribution $P_s(q|m_s)$ of values of $q$ for typical trajectories is Gaussian. We then calculate the mean $q_s$ and the variance $\sigma_s^2$ of this distribution. To calculate the rate function we evaluate 
\bea
\label{big_si}
I(m_s) =-s m_s -q_s  -\frac{K}{2}\sigma_s^2.
\eea
This procedure gives one point $(m_s, I(m_s))$ on the curve $I(m)$. Repeating the procedure for distinct values of $s$ gives the results shown in \f{fig_eden_lattice}(a). In \f{fig_scholes} we show the same data together with the Jensen bound on the rate function, \eqq{jensen} (\eq{big_si} with the term in $\sigma_s^2$ omitted).

\section{Sampling multiple extensive order parameters}
\label{toy}

Consider a ``toy'' reversible model of growth in which blue and red particles add to the system with constant rates $\lb =\lr= c/2$, and depart with constant rates $\gb=\gr = \gamma/2$. We choose $c>\gamma$ so that the system ``grows''. This model can also be regarded as a reversible version of the coin-toss model in which the result (blue $=$ heads or red $=$ tails) of a coin toss can be erased. Define the extensive magnetization $M \equiv b-r$ and system size $N \equiv b+r$, where $b$ and $r$ are the numbers of blue and red particles in the system. Define the intensive variables $\mu \equiv M/K$ and $n \equiv N/K$, where $K$ is the total number of events. It is straightforward to work out the large-deviation rate functions for $\mu$ and $n$. The likelihood that $M$ increases by 1 in a given move is $(c/2+\gamma/2)/(c+\gamma) = 1/2$, and the likelihood that it decreases by 1 is $1/2$. Thus in $K$ moves the likelihood of magnetization $M$ is $P(M,K) = 2^{-K} \binom{K}{(M+K)/2}$. Taking logarithms and using Stirling's formula gives the coin-toss rate function
\beq
\label{rf1}
I_{\rm coin}(\mu)=  \frac{1-\mu}{2} \ln \left(1-\mu\right)+\frac{1+\mu}{2}\ln \left( 1+\mu \right).
\eeq
The likelihood than any move increases $N$ by 1 is $c/(c+\gamma)$. $N$ decreases by 1 with likelihood $\gamma/(c+\gamma)$. Thus $P(N,K) = (\gamma/(c+\gamma))^{K} (c/\gamma)^{(N+K)/2} \binom{K}{(N+K)/2}$. Taking logarithms and using Stirling's formula gives the ``biased coin-toss'' rate function
\beq
\label{rf2}
I_{n_0}(n) =\frac{1-n}{2} \ln{\frac{1-n}{1-n_0}}+\frac{1+n}{2}\ln{\frac{1+n}{1+n_0}},
\eeq
where $n_0 \equiv (c-\gamma)/(c+\gamma)$ is the mean long-time value of $n$. (In the limit $\gamma = 0$ we recover the regular coin-toss model; in this case $N=K$ and $m \equiv M/N$ is equal to $\mu$). 

To obtain the rate function for $m = \mu/n$, we change variables from $(\mu,n)$ to $(m,n)$ and marginalize: 
\beq
I_{\rm toy}(m)= -\frac{1}{K} \ln \int_{-\infty}^\infty |n| {\rm d} n \, \e^{-K(I_{\rm coin}(mn)+I_{n_0}(n))},
\eeq
where $|n|$ is the Jacobian of the transformation. For large $K$ we can evaluate the integral by saddle-point approximation to get
\beq
\label{rfmarginal}
I_{\rm toy}(m) = \min_n \left[ I_{\rm coin}(m n)+I_{n_0}(n) \right].
\eeq
In \f{fig_scholes2} we show a comparison between \eq{rfmarginal} and the form $I_{\rm coin}(m n_0)$ that would obtain if $N$ could not fluctuate. The distinction between these quantities in the tail of the rate function indicates that a rare value of $m$ can be obtained via a rare value of $M$ and a typical value of $N$, or (with greater probability) via a less rare value of $M$ and an atypical value of $N$. 

\begin{figure}[t] 
   \centering
   \includegraphics[width=0.5\linewidth]{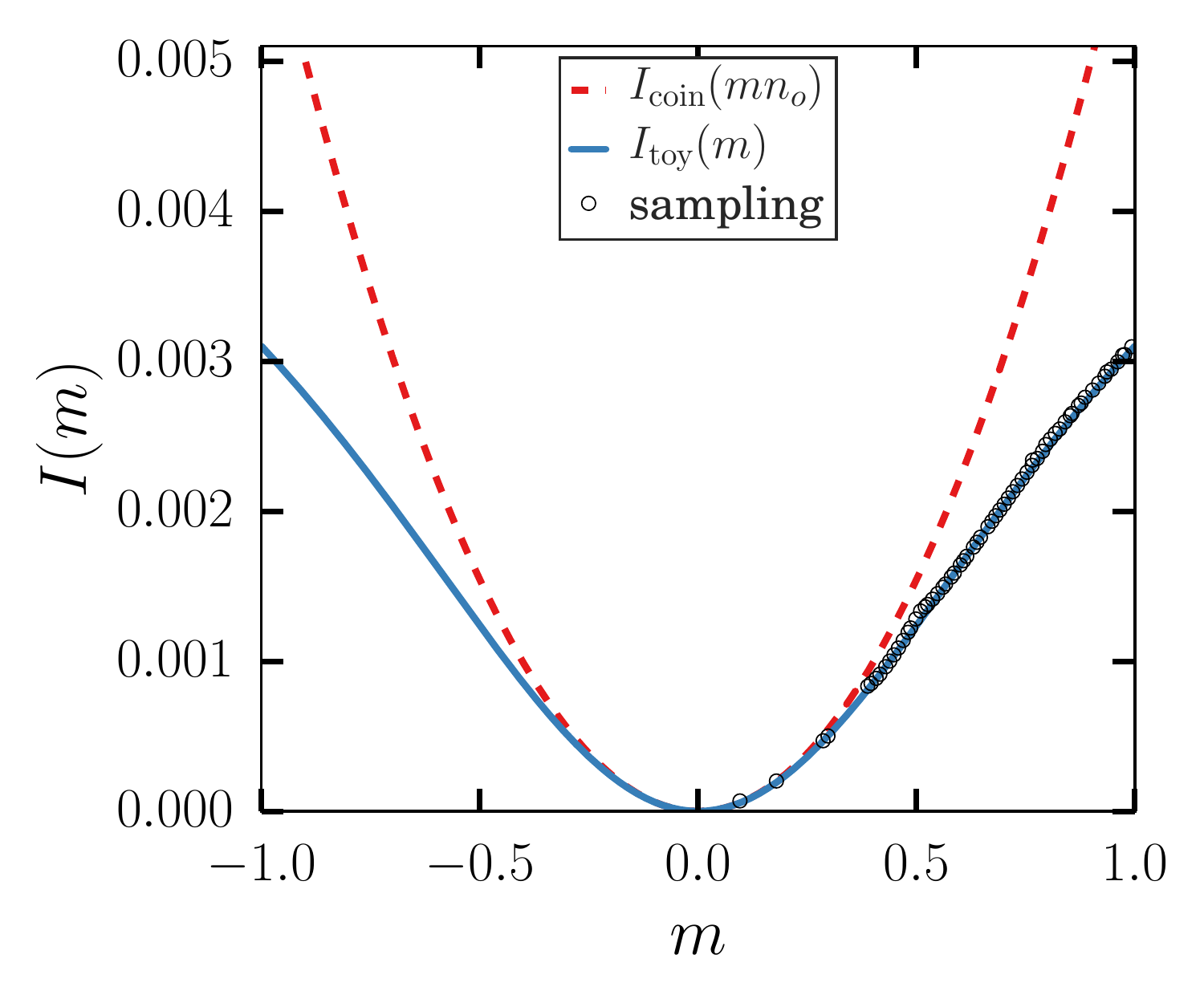} 
   \caption{Rate functions for the magnetization in the ``toy" reversible model (blue) and for the same model under the approximation in which system size cannot fluctuate (red).}
   \label{fig_scholes2}
\end{figure}

These results can also be obtained using the reference-model method, provided that one samples both $M$ and $N$. Equations \eq{ref_model_gen} and \eq{big_gen} provide a means of doing so. We introduce a vector $\bm{s} = (s,s')$ of control parameters, where $s$ biases $M$ and $s'$ biases $N$. The reference model \eq{ref_model_gen} is then $\lambda_{b,\bm{s}}=\lb \e^{-s-s'}$, $\lambda_{r,\bm{s}}=\lr \e^{s-s'}$, $\gamma_{b,\bm{s}}=\gb \e^{s+s'}$, $\gamma_{r,\bm{s}}=\gr \e^{-s+s'}$. The typical values $\mu_{\bm{s}}$ and $n_{\bm{s}}$ of $\mu$ and $n$ are
\bea
\label{sub1}\mu_{\bm{s}} &=& -\tanh s \implies s = -\tanh^{-1} \mu_{\bm{s}};\\
\label{sub2}n_{\bm{s}} &=& \frac{c \e^{-s'} - \lambda \e^{s'}}{c \e^{-s'} + \lambda \e^{s'}} \implies e^{-2 s'} = \frac{\lambda}{c} \frac{1+ n_{\bm{s}}}{1-n_{\bm{s}}}.
\eea
To evaluate \eqq{big_gen} we note that $q$ is constant and equal to 
\bea
\label{sub3}
q_{\bm{s}} = \ln\frac{\lambda_{b,\bm{s}}+\lambda_{r,\bm{s}}+\gamma_{b,\bm{s}}+\gamma_{r,\bm{s}}}{\lb+\lr+\gb+\gr} = \ln \cosh s+ \ln{\frac{c \e^{-s'} + \lambda \e^{-s'}}{c+\lambda}},
\eea
and so $P_{\bm{s}}(q|\bm{a}_{\bm{s}}) = \delta(q_{\bm{s}}-q)$. Then \eq{big_gen} reads, in the limit of large $K$, 
\beq
I(\mu_{\bm{s}},n_{\bm{s}}) = - s \mu_{\bm{s}} + s' n_{\bm{s}} - q_{\bm{s}},
\eeq
with $s$, $s'$, and $q_{\bm{s}}$ given by \eq{sub1}, \eq{sub2}, and \eq{sub3}, respectively. Making these substitutions gives
\beq
I(\mu_{\bm{s}},n_{\bm{s}}) = I(\mu_{\bm{s}})+I_{n_0}(n_{\bm{s}}).
\eeq
Thus the method provides one point on the rate-function curves \eq{rf1} and \eq{rf2}. Repeating the procedure for a range of values of $s$ and $s'$ generates the whole rate-function surface.

\section{Analytic rate function of the reversible model}
\label{add}

In this appendix we compute analytically an upper bound on the rate function of the reversible model. The full rate function, shown in panels (b) and (c) of \f{2_dimensional_sampling}, contains in addition a contribution from fluctuations (of the weight function $q$), and this must be computed numerically.

Typical trajectories of the reference model possess values of $\mu=M/K$ and $n=N/K$ that satisfy
\beq
\label{eqa1}
\mu_{\bm{s}} = \frac{-c \e^{-s'} \sinh s + \e^{s'} \left[- \gamma(m_{\bm{s}}) \e^{s} + \gamma(-m_{\bm{s}}) \e^{-s} \right]}{R(s,s')}
\eeq
and
\beq
\label{eqb1}
n_{\bm{s}} = \frac{c \e^{-s'} \cosh s - \e^{s'} \left[ \gamma(m_{\bm{s}}) \e^{s} + \gamma(-m_{\bm{s}}) \e^{-s} \right]}{R(s,s')},
\eeq
with  $m_{\bm{s}}\equiv \mu_{\bm{s}}/n_{\bm{s}}$, $\gamma(x) \equiv (1+x)\e^{-J x}/2$, and 
\beq
R(s,s') \equiv c \e^{-s'} \cosh s + \e^{s'} \left[ \gamma(m_{\bm{s}}) \e^{s} + \gamma(-m_{\bm{s}}) \e^{-s} \right]. \hspace{0.9cm}
\eeq
Neglecting fluctuations of $q$, \eqq{big_gen} reads
\beq
\label{eqc}
I_{\rm rev}^{(1)}(\mu_{\bm{s}},n_{\bm{s}}) = - s \mu_{\bm{s}}  - s' n_{\bm{s}}-\ln\frac{R(s,s')}{R(0,0)},
\eeq
where we have made use of the fact that the mean value of $q$ is determined by $m$. We use equations \eq{eqa1} and \eq{eqb1} to determine the values of $(\mu_{\bm{s}},n_{\bm{s}})$ associated with the pair $(s,s')$. We then evaluate \eq{eqc} and select the smallest value associated with a particular value of $m_{\bm{s}}$. The result is one point on the rate-function curve $(m_{\bm{s}},I_{\rm rev}(m_{\bm{s}}))$, i.e.
\beq
\label{rev}
I_{\rm rev}^{(1)}(m) = \min_n  [I_{\rm rev}^{(1)}(m n,n)].
\eeq
A convenient approximation to $I_{\rm rev}^{(1)}(m)$ can be obtained from \eq{eqa1} and \eq{eqb1} by assuming that $N$ does not fluctuate. The result is a bound $I^0_{\rm rev}(m) \geq I_{\rm rev}(m)$, where
\bea
I_{\rm rev}^0(m)= - s_0 m \frac{\Gamma_-(s_0)}{\Gamma_+(s_0)}- \ln \frac{\Gamma_+(s_0)}{\Gamma_+(0)}.
\eea
Here $\Gamma_\pm(s) \equiv c \cosh s \pm \cosh(s - J m) \pm m \sinh(s - J m)$, and 
\beq
s_0 \equiv - \tanh^{-1}\frac{c m - \sinh(J m) + m^2 \sinh(J m)}{
 c + \cosh(J m) - m^2 \cosh(J m)}.
 \eeq
 In the main text we use numerical simulations of the reference model to compute the reversible model's rate function $I_{\rm rev}(m)$ exactly (the above expressions neglect certain fluctuations of the reference-model trajectory ensemble); comparison of numerics and the bound described above, in e.g. \f{2_dimensional_sampling}(c), shows the bound to be reasonably tight.

\end{document}